%% file: NSBH_Model.tex
\documentclass[%
 reprint,
superscriptaddress,
nofootinbib,
 amsmath,amssymb,
 twocolumn,
 aps,
]{revtex4}

\usepackage{graphicx}
\usepackage{dcolumn}
\usepackage{bm}
\usepackage{xcolor}
\usepackage[normalem]{ulem}

\include{macros}

\begin{document}

\preprint{APS/123-QED}

\title{An aligned-spin neutron-star--black-hole waveform model based on the effective-one-body approach and 
numerical-relativity simulations}

\author{Andrew Matas}
\affiliation{Max Planck Institute for Gravitational Physics (Albert Einstein Institute), D-14476 Potsdam, Germany}
\author{Tim Dietrich}
\affiliation{Nikhef, Science Park 105, 1098 XG Amsterdam, The Netherlands}
\affiliation{Institute for Physics and Astronomy, University of Potsdam, Karl-Liebknecht-Str. 24/25, 14776 Potsdam, Germany}
\author{Alessandra Buonanno} 
\affiliation{Max Planck Institute for Gravitational Physics (Albert Einstein Institute), D-14476 Potsdam, Germany}
\affiliation{Department of Physics, University of Maryland, College Park, MD 20742, USA}
\author{Tanja Hinderer}
\affiliation{GRAPPA, Anton Pannekoek Institute for Astronomy and Institute of High-Energy Physics, University of Amsterdam, Science Park 904, 1098 XH Amsterdam, The Netherlands}
\affiliation{Delta Institute for Theoretical Physics, Science Park 904, 1090 GL Amsterdam, The Netherlands}
\author{Michael P\"urrer}
\affiliation{Max Planck Institute for Gravitational Physics (Albert Einstein Institute), D-14476 Potsdam, Germany}
\author{  Francois Foucart}
\affiliation{Department of Physics \& Astronomy, University of New Hampshire, 9 Library Way, Durham NH 03824, USA }
\author{Michael Boyle}
\affiliation{Cornell Center for Astrophysics and Planetary Science, Cornell University, Ithaca, New York 14853, USA}
\author{Matthew D. Duez}
\affiliation{Department of Physics \& Astronomy, Washington State University, Pullman, Washington 99164, USA}
\author{Lawrence E. Kidder}
\affiliation{ Cornell Center for Astrophysics and Planetary Science, Cornell University, Ithaca, New York 14853, USA}
\author{Harald P. Pfeiffer}
\affiliation{Max Planck Institute for Gravitational Physics (Albert Einstein Institute), D-14476 Potsdam, Germany}
\author{Mark A. Scheel}
\affiliation{TAPIR, Walter Burke Institute for Theoretical Physics, MC 350-17, California Institute of Technology, Pasadena, California 91125, USA}

\date{\today}

\begin{abstract}
  After the discovery of gravitational waves from binary black holes (BBHs) and binary neutron stars (BNSs) 
with the LIGO and Virgo detectors, neutron-star--black-holes (NSBHs) are
  the natural next class of binary systems to be observed. In this work, we develop a waveform model 
for aligned-spin neutron-star--black-holes (NSBHs) combining a BBH baseline 
waveform (available in the effective-one-body approach) with a phenomenological description of tidal 
effects (extracted from numerical-relativity simulations), and correcting the amplitude during the late 
inspiral, merger and ringdown to account for the NS tidal disruption. In particular, we calibrate the amplitude 
corrections using NSBH waveforms obtained with the numerical-relativity spectral Einstein code (SpEC) and the 
SACRA code. The model was calibrated using simulations with NS masses in the range $1.2-1.4 M_\odot$, tidal deformabilities up to $4200$ (for a 1.2 $M_\odot$ NS), and dimensionless BH spin magnitude up to 0.9. Based on the simulations used, and on checking that sensible waveforms are produced, we recommend our model to be employed with NS mass in the range  
$1\mbox{--}3 M_\odot$, tidal deformability $0\mbox{--}5000$, and (dimensionless) BH spin magnitude up to $0.9$. We 
also validate our model against two new, highly accurate NSBH waveforms with BH spin 0.9 and mass ratios 3 and 4, 
characterized by tidal disruption, produced with SpEC, and find very good agreement. Furthermore, we compute the unfaithfulness between waveforms from NSBH, BBH, and BNS
  systems, finding that it will be challenging for the advanced
  LIGO-Virgo--detector network at design sensitivity to distinguish different source classes. We perform a Bayesian
  parameter-estimation analysis on a synthetic numerical-relativity signal in zero noise 
to study parameter biases. Finally, we reanalyze GW170817,
    with the hypothesis that it is a NSBH. We do not find evidence to
    distinguish the BNS and NSBH hypotheses, however the posterior for
    the mass ratio is shifted to less equal masses under the NSBH
    hypothesis.
\end{abstract}

\maketitle

\section{Introduction}

In their first two observing runs (O1 and O2), Advanced LIGO
\cite{TheLIGOScientific:2014jea} and Advanced Virgo
\cite{TheVirgo:2014hva} have observed gravitational
waves (GWs) from ten binary black holes (BBHs) and one binary neutron
star (BNS), GW170817 \cite{LIGOScientific:2018mvr}.  Recently, in the
third observing run (O3), a second BNS, GW190425, was discovered 
\cite{Abbott:2020uma}. Other groups have reported additional GW observations 
analyzing the public data from the first two runs
\cite{Venumadhav:2019lyq,Zackay:2019tzo,Zackay:2019btq}. Neutron-star--black-holes (NSBHs)
may be the next source class to be discovered. Given the lack of a
detection in O1 and O2, the rate of NSBHs is uncertain. However,
based on estimates from Ref.~\cite{Abadie:2010cf}, the expected number of
NSBH detections is $0_{-0}^{+19}$ O3 and $1_{-1}^{+91}$ in O4
\cite{Aasi:2013wya}, where the central value is the median 
and the error bars give the 90\% credible interval. As of this writing, in O3, the LIGO and Virgo
Collaborations have published seven circulars via the Gamma-ray
Coordinates Network (GCN) describing detection candidates for which
the probability of the system being a NSBH is larger than 1\%, and
for which the candidate has not been retracted
\cite{GCN25324,GCN25549,GCN25695,GCN25814,GCN25876,GCN26350,GCN26640}. Furthermore,
GW data alone does not exclude the possibility that
GW170817 is a NSBH~\cite{Hinderer:2018pei,Coughlin:2019kqf,LIGOScientific:2019eut}, and
it has also been suggested that GW190425 could be a NSBH
\cite{Han:2020qmn,Kyutoku:2020xka}. Therefore it is timely to develop
methods that can be used to study NSBHs in GW data.

NSBH binaries exhibit a rich phenomenology that is imprinted on the
gravitational waveform (for a review see Ref.~\cite{Shibata:2011jka}). First, as is the case for BNS systems, finite
size effects cause a dephasing of the waveform relative to a BBH with
the same masses and spins \cite{Flanagan:1997fn,Flanagan:2007ix,Vines:2010ca,Pannarale:2011pk}. Additionally, the
amplitude of NSBH waveforms can be affected by tidal forces~\cite{Kyutoku:2010zd}. For
unequal mass ratios and slowly spinning BHs, the
amplitude of the waveform is well-described by a BBH
\cite{Foucart:2013psa}. On the other hand, for near-equal mass ratios
or for highly spinning BHs, depending on the NS equation of state (EOS),  
the NS can undergo tidal disruption, in which the star is ripped apart as it approaches 
the BH~\cite{Foucart:2010eq,Kyutoku:2010zd,Kyutoku:2011vz,Foucart:2012vn,Kawaguchi:2015bwa}. If the disruption takes place 
before the NS crosses the innermost stable circular orbit, then the material ejected from 
the NS can form a disk around the BH~\cite{Pannarale:2010vs,Foucart:2012nc,Lovelace:2013vma}. If so, 
starting at a characteristic (cutoff) frequency~\cite{Kyutoku:2010zd,Kawaguchi:2017twr,Pannarale:2015jia}, 
the amplitude of the waveform is strongly suppressed, and the ringdown stage is reduced or even effaced. 
The details of this process contain information about the NS EOS. 
Additionally, NSBH mergers can be the progenitors of short gamma-ray bursts \cite{Paczynski:1986px,Eichler:1989ve,Paczynski:1991aq,Meszaros:1992ps,Narayan:1992iy,Nakar:2007yr}, and the disk around the remnant BH and dynamical ejecta can provide the engine for the kilonova 
signal~\cite{Metzger:2016pju,Tanaka:2016sbx}, like the ones observed for GW170817~\cite{LIGOScientific:2018mvr,GBM:2017lvd}.

In order to take advantage of this potentially rich source of
information, it is crucial to have a fast and accurate waveform model
capturing effects due to relativistic matter, which can be used in
analyzing GW data. Several approaches exist for describing finite-size
effects in BNS systems. Tidal corrections~\cite{Flanagan:1997fn,Flanagan:2007ix,Vines:2010ca,
PhysRevD.80.084035,PhysRevD.80.084018} have been incorporated in 
the effective one-body (EOB) formalism~\cite{PhysRevD.59.084006,PhysRevD.62.064015,Damour:2000we} 
in Refs.~\cite{Damour:2009wj,Damour:2012yf,Steinhoff:2016rfi,Hinderer:2016eia,Nagar:2018zoe,Lackey:2018zvw}. 
References~\cite{Dietrich:2017aum,Dietrich:2019kaq} developed a flexible
technique that starts from a point-mass BBH baseline waveform, and applies tidal-phase modifications 
by fitting a Pad\'e-resummed post-Newtonian (PN)--based ansatz to the phasing extracted from numerical-relativity (NR) 
simulations (henceforth, we refer to this as the NRTidal approach). These corrections have been applied to BBH baselines produced
within the EOBNR framework \cite{Bohe:2016gbl}, and within the 
inspiral-merger-ringdown phenomenological (IMRPhenom) approach \cite{Khan:2015jqa,Hannam:2013oca}.

\begin{table}[]
    \centering
    \begin{tabular}{c|c|c}
\hline
\hline          
Name in this paper & LAL name & Ref.  \\
          \hline
         SEOBNR\_BBH &   \texttt{SEOBNRv4\_ROM}  & \cite{Bohe:2016gbl} \\
         SEOBNR\_BNS & \texttt{SEOBNRv4\_ROM\_NRTidalv2} & \cite{Dietrich:2019kaq} \\
        SEOBNR\_NSBH & \texttt{SEOBNRv4\_ROM\_NRTidalv2\_NSBH} &  this paper \\
\hline
    \end{tabular}
    \caption{Dictionary relating the names we use in this paper for several waveforms 
from the SEOBNR family and the corresponding names of the waveforms implemented in the LIGO Algorithms Library (LAL). 
The second and third waveforms use tidal effects within the NRTidal approach.}
    \label{tab:lal-names}
\end{table}

There have been several previous works constructing NSBH waveforms. An
aligned-spin NSBH waveform model was developed in
Refs.~\cite{Lackey:2011vz,Lackey:2013axa}, but it covered a limited range of mass
ratios. In Ref.~\cite{Kumar:2016zlj}, this waveform model was used in
parameter and population studies in conjunction with a former 
version of the EOBNR BBH baseline~\cite{Taracchini:2013rva}. A 
NSBH model called PhenomNSBH, 
which was constructed using a similar approach to modeling NSBHs
as the one discussed in this paper
 but developed within the 
IMRPhenom approach,  was recently 
put forward in Ref.~\cite{Thompson:2020nei}. This model 
uses the method of Ref.~\cite{Pannarale:2015jka} to describe
tidal disruption of the amplitude, 
and uses the tidal phase corrections from Ref.~\cite{Dietrich:2019kaq}.

In this work we develop a frequency-domain model for the dominant, quadrupolar
multipole of GWs emitted by aligned-spin NSBH systems. Together with 
the recent waveform model of Ref.~\cite{Thompson:2020nei}, these are the first NSBH models covering a
wide range of mass ratios and spin that can be used to analyze GW
data. In this paper, we refer to our model as \nsbh, which has 
already been implemented in the LIGO Algorithms Library (LAL) \cite{lalsuite}. 
In Table~\ref{tab:lal-names} we provide a dictionary between the names
we use in this work, and the name as implemented in LAL. The amplitude is based on an EOBNR BBH
baseline model that we refer to as \bbh~\cite{Bohe:2016gbl}. We apply corrections
inspired by Pannarale \emph{et al.} \cite{Pannarale:2015jka} to
account for tidal disruption. We have adapted the corrections of Ref.~\cite{Pannarale:2015jka}, 
originally developed for a former version of the IMRPhenom BBH  model \cite{Santamaria:2010yb}, for use 
with EOBNR waveforms~\cite{Bohe:2016gbl}, augmented with reduced-order modeling 
(ROM) \cite{Purrer:2014fza,Purrer:2015tud} to enhance the speed. Differently from 
Ref.~\cite{Thompson:2020nei}, which uses the fit from  Pannarale \emph{et al.} \cite{Pannarale:2015jka}, 
here we have performed a fit incorporating results from the new NSBH simulations at our disposal as described in Sec.~\ref{sec:NR-data}. While the publicly available SXS simulations were not used for calibration of Ref.~\cite{Thompson:2020nei}, these waveforms were used for validation and good agreement was found. The phase is
computed by applying tidal corrections to the EOBNR BBH baseline~\cite{Bohe:2016gbl} 
using the NRTidal approach, as in
\bns~\cite{Dietrich:2019kaq}.  As shown in Ref.~\cite{Foucart:2018lhe}, even though the tidal
corrections from \bns\ were derived from BNS simulations, they give
good agreement with NSBH simulations. 

The rest of this paper is organized as follows. In Sec.~\ref{sec:Construction}, we describe the construction of the
waveform model. We review properties of NSBH systems in
Sec.~\ref{sec:nsbh-properties}, summarize the NR waveforms that we use in Sec.~\ref{sec:NR-data}, give an outline of the waveform
model in Sec.~\ref{sec:parameterization}, summarize the procedure
used to calibrate the amplitude correction, assess their accuracy 
by computing the unfaithfulness, and compare the NSBH waveforms 
to NR simulations in Sec.~\ref{sec:fits}. Then, we discuss the regime of validity in
Sec.~\ref{sec:validity}. Next, we apply the waveform model to several data
analysis problems. First, in Sec.~\ref{sec:fitting-factor} we
estimate when the advanced LIGO-Virgo detector
network at design sensitivity can distinguish NSBH and BBH, and NSBH and BNS, systems. Then,
in Sec.~\ref{sec:SXS-PE} we perform parameter-estimation Bayesian analysis on an
NR-waveform, hybridized to an analytical waveform at low frequency, and show the differences 
between recovering this waveform with a BBH and NSBH model. Finally, we
  reanalyze GW170817 under the hypothesis that it is a NSBH in
  Sec.~\ref{sec:GW170817}. We conclude in
Sec.~\ref{sec:Conclusions} by summarizing the main points and lay
out directions for future improvements. Finally, in Appendix~\ref{app:w_definition}, 
we give explicit expressions defining the waveform model.

We work in units with $G=c=1$. The symbol
$M_\odot$ refers to the mass of the sun.

\section{Constructing the NSBH waveform model}
\label{sec:Construction}

\begin{figure*}[ht]
    \centering
    \includegraphics[width=\textwidth]{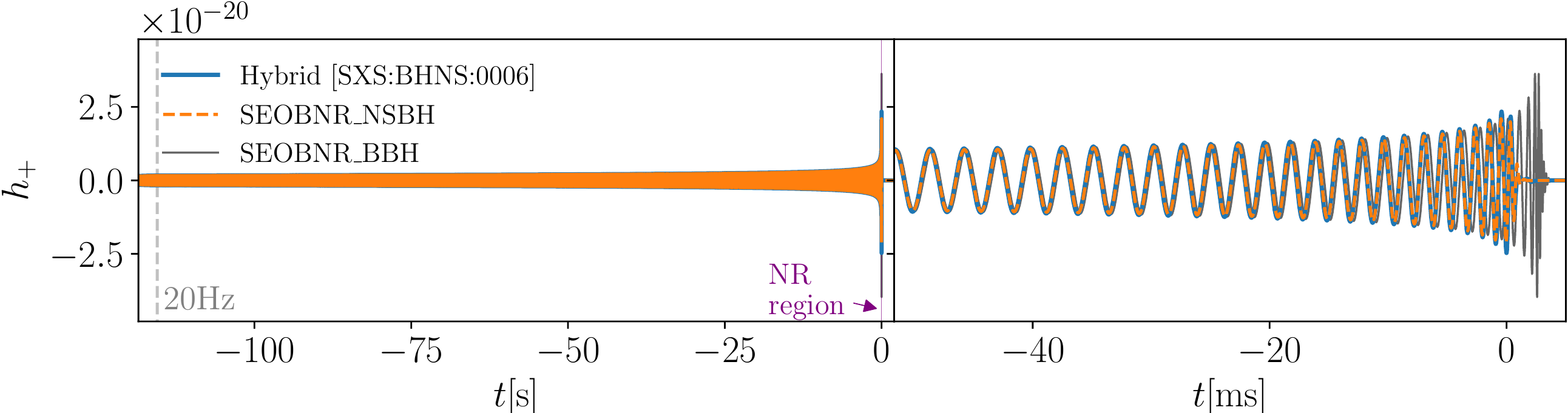}
    \caption{We compare an NR-hybrid waveform, which is constructed by stitching together the \texttt{SXS:BHNS:0006} 
($M_{\rm NS} = 1.4 M_\odot$, $M_{\rm BH} = 2.1 M_\odot$, $\Lambda_{\rm NS}=791$) 
and \bns\ ($m_1= 1.4 M_\odot$, $m_2 = 2.1 M_\odot$, $\Lambda_1=0$, $\Lambda_2=791$) waveforms, 
with the \nsbh\ ($M_{\rm NS} = 1.4 M_\odot$, $M_{\rm BH} = 2.1 M_\odot$, $\Lambda_{\rm NS}=791$) and 
\bbh\ ($m_1= 1.4 M_\odot$, $m_2 = 2.1 M_\odot$, $\Lambda_1=\Lambda_2=0$) waveforms.  The strain is produced by a source at a distance of 1 Mpc. The definition of the tidal parameter $\Lambda_{\rm NS}$ is given in Eq.~\ref{eq:lambda-def}.
In the left panel, we show the waveforms during the long inspiral and mark $20\, {\rm Hz}$, which is the lower frequency
      typically used for LIGO-Virgo parameter-estimation analyses. We also indicate the region toward merger where 
      the NR data are available. 
      In the right panel, we zoom into the last stages of the inspiral and merger. Due to tides, the hybrid waveform and
      \nsbh\ have a faster inspiral, and end earlier than the
      corresponding point-mass \bbh\ waveform. Furthermore,
      because of tidal disruption, the hybrid and \nsbh\ waveforms have no ringdown phase, but end abruptly when
      the NS gets disrupted. Overall, we find very good agreement between
      \nsbh\ and NR-hybrid waveforms throughout the entire coalescence.}
\label{fig:illustration}
\end{figure*}

\subsection{NSBH binary properties}
\label{sec:nsbh-properties}

We begin by providing a brief description of the final stages of a NSBH coalescence,  
identifying the main features of the process and the physical properties of the 
remnant BH. Here, we mainly follow the discussion in Refs.~\cite{Shibata:2011jka,Foucart:2012nc,Pannarale:2015jka}. 

The two bodies spiral in due to the loss of energy from the emission of GWs. If the NS approaches close enough to the BH, tidal forces exerted by the BH on the NS  can overcome the self gravity of the NS, causing the star to loose mass. This process is called \emph{mass shedding}. This in turn often leads to \emph{tidal disruption}, in which the NS is completely torn apart by the strong gravitational field of the BH.
Let us denote with $r_{\rm tide}$ the separation of the binary at which mass shedding begins.  
To understand the fate of the NS and 
the characteristics of the GW signal emitted during the last stages of inspiral, plunge 
and merger, we compare $r_{\rm tide}$ to the location of the innermost-stable 
circular orbit (ISCO) (which marks the beginning of the plunge).
If $r_{\rm tide}<  r_{\rm ISCO}$, the NS is swallowed by the BH, without loss of material. By contrast, if $r_{\rm tide} >  r_{\rm ISCO}$, mass is ejected from the NS 
before it plunges. If the NS is far away from the ISCO when it is disrupted, matter may form an accretion
disk (torus) around the BH after merger. It has been shown (e.g., see Refs~\cite{Kyutoku:2010zd,Kyutoku:2011vz,Foucart:2012nc,Lovelace:2013vma} and also below) that 
the disruption affects the GW signal for NSBH binaries with either nearly equal masses 
or large BH spins aligned with the orbital angular  momentum, because for 
those systems the condition $r_{\rm tide} <  r_{\rm ISCO}$ is satisfied. In Fig.~\ref{fig:illustration}, we show an illustration of the effect of tidal disruption on the GW waveform for an example NR hybrid with mass ratio 1.5.

Let us now estimate the radial separation at which mass shedding occurs,
$r_{\rm tide}$, by imposing that the tidal force from the BH balances 
the self-gravity of the NS. As described in Ref.~\cite{Foucart:2012nc}, 
in the Newtonian limit, $r_{\rm tide}$ can be estimated as $r_{\rm tide} 
\approx \xi_{\rm Newt} R_{\rm NS}$, where  $\xi_{\rm Newt}=(3Q)^{1/3}$, $R_{\rm NS}$ 
is the NS radius in isolation, $Q\equiv M_{\rm BH}/M_{\rm NS}$ is the mass ratio, $M_{\rm BH}$ is the mass of the BH, and $M_{\rm NS}$ is the mass of the NS. 
The factor of 3 is an estimate obtained by matching with NR simulations. 
This estimate can be improved by accounting for relativistic effects due to the
large compactness of the NS, $C_{\rm NS}\equiv M_{\rm NS}/R_{\rm NS}$
\cite{Foucart:2012nc}. First, $r_{\rm tide}$ is reduced by a factor
$(1-2C_{\rm NS})$ relative to the Newtonian estimate; this factor
enforces the absence of tidal disruption in the BH limit $C_{\rm
  NS}\rightarrow 1/2$. Second, point-mass motion in the Kerr
metric leads to a correction factor $\xi$ which differs from the
Newtonian estimate~\cite{Foucart:2012nc,1973ApJ...185...43F}. 
Combining these effects, we have
\begin{equation}
    r_{\rm tide} = \xi (1-2C_{\rm NS})R_{\rm NS}\,.
\end{equation}
The relativistic correction parameter $\xi$ is determined by solving the algebraic equation (we take the largest positive root of this equation)
\begin{equation}
   \left(\frac{\xi}{\xi_{\rm Newt}} \right)^3 = \frac{\xi^2 - 2 Q C_{\rm NS} \xi + Q^2 C_{\rm NS}^2 \chi_{\rm BH}^2}{\xi^2 - 3 QC_{\rm NS} \xi + 2 \chi_{\rm BH} \sqrt{Q^3 C_{\rm NS}^3 \xi}}, 
\end{equation}
where $\chi_{\rm BH}$ is the spin of the BH. We can associate to the tidal-disruption separation a frequency, which is more useful 
in the context of modeling the gravitational waveform, as follows
\begin{equation}
    f_{\rm tide} = \frac{1}{\pi(\chi_{\rm BH} M_{\rm BH} + \sqrt{r_{\rm tide}^3/M_{\rm BH}})},
\end{equation}
which is obtained from the (circular orbit) relation between radial separation and (angular) orbital frequency in the Kerr geometry.\footnote{Note that in \cite{Pannarale:2015jka}, the formula for $f_{\rm tide}$ is written in terms of the final, rather than initial, BH mass and spin. In LAL, \nsbh\ is implemented with the final BH properties. We became aware of this point during a late stage of this work. The fits in this work were done self-consistently using the final BH properties. We have checked that when we replace $M_{f}$ and $\chi_f$ by $M_{\rm BH}$ and $\chi_{\rm BH}$ in the expression for $f_{\rm tide}$, mismatches with \nsbh\ are $\mathcal{O}(10^{-4})$ or less across parameter space. We thank the internal LIGO review team for bringing this to our attention.}

The NS compactness $C_{\rm NS}$, which depends on the NS EOS,  enters the expression for $r_{\rm tide}$. In order to avoid making an assumption about the EOS, it is more convenient to work in terms of the
dimensionless tidal-deformability parameter $\Lambda_{\rm NS}$, which relates
the quadrupole moment of the NS to the tidal field of the companion. The tidal parameter is
determined by the compactness of the NS and the tidal Love number
$k_2$ as follows
\begin{equation}
\label{eq:lambda-def}
    \Lambda_{\rm NS} = \frac{2}{3} \frac{k_2}{C_{\mathrm{NS}}^5}.
\end{equation}
We take the tidal parameter of the BH to be zero. The tidal parameter of non-spinning BHs was shown to be zero in  Ref.~\cite{Binnington:2009bb} f.
We can relate $\Lambda_{\rm NS}$ and $C_{\rm NS}$ in an equation-of-state independent way with the $\Lambda_{\rm NS}-C_{\rm NS}$ relation \cite{Yagi:2016bkt}
\begin{equation}
    C_{\rm NS} = \sum_{k=0}^2 a_k (\ln \Lambda_{\rm NS})^k,
\end{equation}
with $a_0=0.360, a_1=-0.0355, a_2=0.000705$. In order to achieve
continuity with BBH waveforms in the limit $\Lambda_{\rm
  NS}\rightarrow 0$, for $\Lambda_{\rm NS}\leq1$, we replace the
$\Lambda_{\rm NS}-C$ relation with a cubic polynomial which interpolates from
$\Lambda_{\rm NS}=1$ to $C_{\rm NS}=1/2$ at $\Lambda_{\rm NS}=0$, and
it is continuous and once differentiable at $\Lambda_{\rm NS}=1$.
The universal relations are also used in Ref.~\cite{Thompson:2020nei}.

The matter ejected from the NS, during tidal disruption, can
remain bound, forming a disk (torus) around the remnant BH. The mass of this
remnant torus, $M_{\rm{b,torus}}$, can be determined in terms of
the baryonic mass of the NS using fits from Ref.~\cite{Foucart:2012nc} (see
also more recent simulations performed in Ref.~\cite{Foucart:2019bxj})
\begin{equation}
    \frac{M_{\rm b,torus}}{M_{\rm b,NS}} = {\rm max} \left( \frac{0.296 r_{\rm tide} - 0.171 r_{\rm ISCO}}{R_{\rm NS}}, 0 \right),
\end{equation}
where the ISCO radius ($r_{\rm ISCO}$) in the Kerr spacetime is given by
\begin{subequations}
\begin{eqnarray}
    r_{\rm ISCO} &=& 3+Z_2 \mp \sqrt{(3-Z_1)(3+Z_1+2Z_2)}\,, \\
    Z_1 &=& 1+(1-\chi_{\rm BH}^2)^{1/3}\left[(1+\chi_{\rm BH})^{1/3}+ \right . \nonumber \\
&& \left . (1-\chi_{\rm BH})^{1/3}\right]\,,\\
    Z_2 &=& \sqrt{3\chi_{\rm BH}^2 + Z_1^2},
\end{eqnarray}
\end{subequations}
where the $\mp$ sign holds for prograde (retrograde) orbits. 

As mentioned above, the onset of mass shedding occurs when the objects
approach within a distance $r_{\rm tide}$ before the NS cross the
ISCO. However, the ISCO does not introduce a definite feature in the
gravitational waveform. In order to identify the onset of tidal
disruption with a definite feature in an NR waveform, in our model we
compare $f_{\rm tide}$ to the ringdown frequency of the final BH,
$f_{\rm RD}$, which is the frequency of least-damped quasinormal mode
of the final BH. The ringdown frequency can be computed from the final
mass and spin using fitting formulas from Ref.~\cite{Berti:2005ys}. To
obtain the final mass and spin from the initial parameters of the
binary, we use the fits performed by Ref.~\cite{Zappa:2019ntl}, 
which account for the ejected mass.

\subsection{Numerical-relativity waveforms}
\label{sec:NR-data}

In this section we briefly describe the NR data used to construct and
validate the model. The Simulating eXtreme Spacetimes (SXS)
collaboration has publicly released data from seven simulations described
in Refs.~\cite{PhysRevD.88.064017,Foucart:2018lhe}, which were produced
using the Spectral Einstein Code (\verb+SpEC+), see Ref.~\cite{SPECWebsite}. The hyrodynamical part of the code is described in Ref.~\cite{PhysRevD.78.104015,PhysRevD.87.084006}. These
configurations do not contain spinning BHs, but do include mergers
with and without tidal disruption. These simulations use an ideal gas EOS with polytropic index $\Gamma=2$, 
except for the mass ratio 3 simulation
\texttt{SXS:BHNS:0003}, which uses a piecewise polytropic ansatz calibrated to the H1 EOS, see Ref.~\cite{Read:2008iy}. We refer the reader to Ref.~\cite{Foucart:2018lhe} for further explanation.
For five of these
simulations the NS spin is zero, and we use these simulations to
fit the model as described in Sec.~\ref{sec:fits}. We use the other
two simulations for verification.
Additionally, \verb+SpEC+ has simulated nine systems with large BH spin in Ref.~\cite{Foucart:2014nda}, 
using the more advanced temperature and composition dependent LS220 EOS \cite{Lattimer:1991nc}, which we
also use to fit our waveform model. Finally, we validate our NSBH model also against 
two new SXS waveforms, \texttt{Q3S9} and \texttt{Q4S9}, which are highly accurate
simulations describing disruptive mergers with large BH spin. These simulations were also performed using the $\Gamma=2$ EOS. We give the parameters of all SXS
waveforms used here in Table~\ref{tab:SXS-spinning-BH}. 

In fitting the model, we also use 134 simulations of irrotational NSs
performed with the \verb+SACRA+ code \cite{Yamamoto:2008js}, which were
presented in Refs.~\cite{Kyutoku:2010zd,Kyutoku:2011vz}. These simulations span
the mass ratios $Q=\{2,3,4,5\}$, BH spins $\chi_{\rm BH}=\{-0.5,0,0.25,0.5,0.75\}$,  
and a range of piecewise polytropic EOS. The parameters for all of the waveforms
and the EOS used are given Table II of Ref.~\cite{Lackey:2013axa}. Whereas the large number of \verb+SACRA+  
waveforms lets us probe a wide parameter range, these waveforms are shorter and 
of lower accuracy than the publicly available  \verb+SpEC+ waveforms as well as  \texttt{Q3S9} and \texttt{Q4S9}, due to finite numerical resolution
and non-negligible eccentricity in the initial data. We note that these simulations predate the public SXS simulations by a number of years.

\begin{table*}[ht]
    \centering
\begin{tabular}{c|c|c|c|c|c|c|c|c}
\hline
\hline
   Label  & $\frac{M_{\rm BH}}{M_\odot}$ & $\frac{M_{\rm NS}}{M_\odot}$  & $Q$ & $\chi_{\rm BH}$ & $\chi_{\rm NS}$ & $\Lambda_{\rm NS}$ & $N_{\rm GW}$  & $e_{\rm res}$ \\
\hline
  \texttt{SXS:BHNS:0001}  & 8.4 & 1.4 & 6 &  0 & 0 & 526 & 25.3 &$ <1.3 \times 10^{-3}$  \\
  \texttt{SXS:BHNS:0002}  & 2.8 & 1.4 & 2 & 0  & 0 & 791 & 26.1 & $ < 5 \times 10^{-4}$ \\
  \texttt{SXS:BHNS:0003}  & 4.05 & 1.35 & 3 &  0 &  0  &  624& 12.3 & $  7.9 \times 10^{-3} $ \\
  \texttt{SXS:BHNS:0004}  & 1.4 & 1.4 & 1 & 0 & 0  & 791 & 24.5 & $ <  6.0 \times 10^{-5}$ \\
  \texttt{SXS:BHNS:0006}  & 2.1 & 1.4 & 1.5 & 0 & 0 & 791 & 33.2 & $<  2.7 \times 10^{-4}$ \\  
    \texttt{M12-7-S8-LS220} & 7 & 1.2 & 5.8  & 0.8 & 0 & 1439 & 17.8 & $2.7 \times 10^{-2}$ \\
    \texttt{M12-7-S9-LS220} & 7 & 1.2 & 5.8  & 0.9 &  0 & 1439 & 18.9 & $2.6 \times 10^{-2}$ \\
    \texttt{M12-10-S8-LS220} & 10 & 1.2 & 8.3 & 0.8 &  0 & 1439 & 20.3 & $3.1 \times 10^{-2}$ \\
    \texttt{M12-10-S9-LS220} & 10  & 1.2 & 8.3  & 0.9 &  0 & 1439 & 22.1 & $3.3 \times 10^{-2}$ \\
    \texttt{M14-7-S7-LS220} & 7 & 1.4 & 5 & 0.7 &  0 & 536 & 10.6 & $3.9 \times 10^{-2}$ \\
    \texttt{M14-7-S8-LS220} & 7 & 1.4 & 5 & 0.8 &  0 & 536 & 11.7 & $3.7 \times 10^{-2}$\\
    \texttt{M14-7-S9-LS220} & 7 & 1.4  & 5& 0.9 &  0 & 536 & 12.5 & $3.7 \times 10^{-2}$\\
    \texttt{M14-10-S8-LS220} & 10 & 1.4 & 7.1  & 0.8 &  0 & 536 & 15.1 & $4.2 \times 10^{-2}$ \\
    \texttt{M14-10-S9-LS220} & 10 & 1.4 & 7.1 & 0.9 &  0 & 536 & 16.8 & $4.3 \times 10^{-2}$ \\
     \hline
   \texttt{SXS:BHNS:0005}  & 1.4 & 1.4 & 1 & 0 &  -0.2   &  791  & 21.6 & $5.0 \times 10^{-4}$ \\
  \texttt{SXS:BHNS:0007}  & 2.8 & 1.4 & 2 & 0 &  -0.2 & 791   & 24.7 & $<4.7 \times 10^{-4}$ \\   
    \texttt{Q3S9} & 4.2 & 1.4 & 3 & 0.9 & 0 & 791 & 26.5 & $5.4 \times 10^{-4}$ \\
    \texttt{Q4S9} & 5.6 & 1.4 & 4 & 0.9 & 0 & 791 & 31.4  & $1.7 \times 10^{-3}$ \\
   \hline
       \label{tab:SXS-spinning-BH}
\end{tabular}
    \caption{Parameters for the SXS NSBH waveforms used in this work. The simulations above the horizontal line were used to fit the NSBH model, the simulations below the line are used for validation. Parameters for the other waveforms that we employ to fit the model were produced by the {\tt SACRA} code, and are given in Table II of Ref.~\cite{Lackey:2013axa}. We also report the number of GW cycles, $N_{\rm GW}$, computed up to the peak of the dominant GW mode. Finally we display the residual eccentricity, $e_{\rm res}$. }
\end{table*}

\subsection{Parameterization of the NSBH waveform model}
\label{sec:parameterization}

\begin{figure*}[ht]
    \centering
    \includegraphics[width=\textwidth]{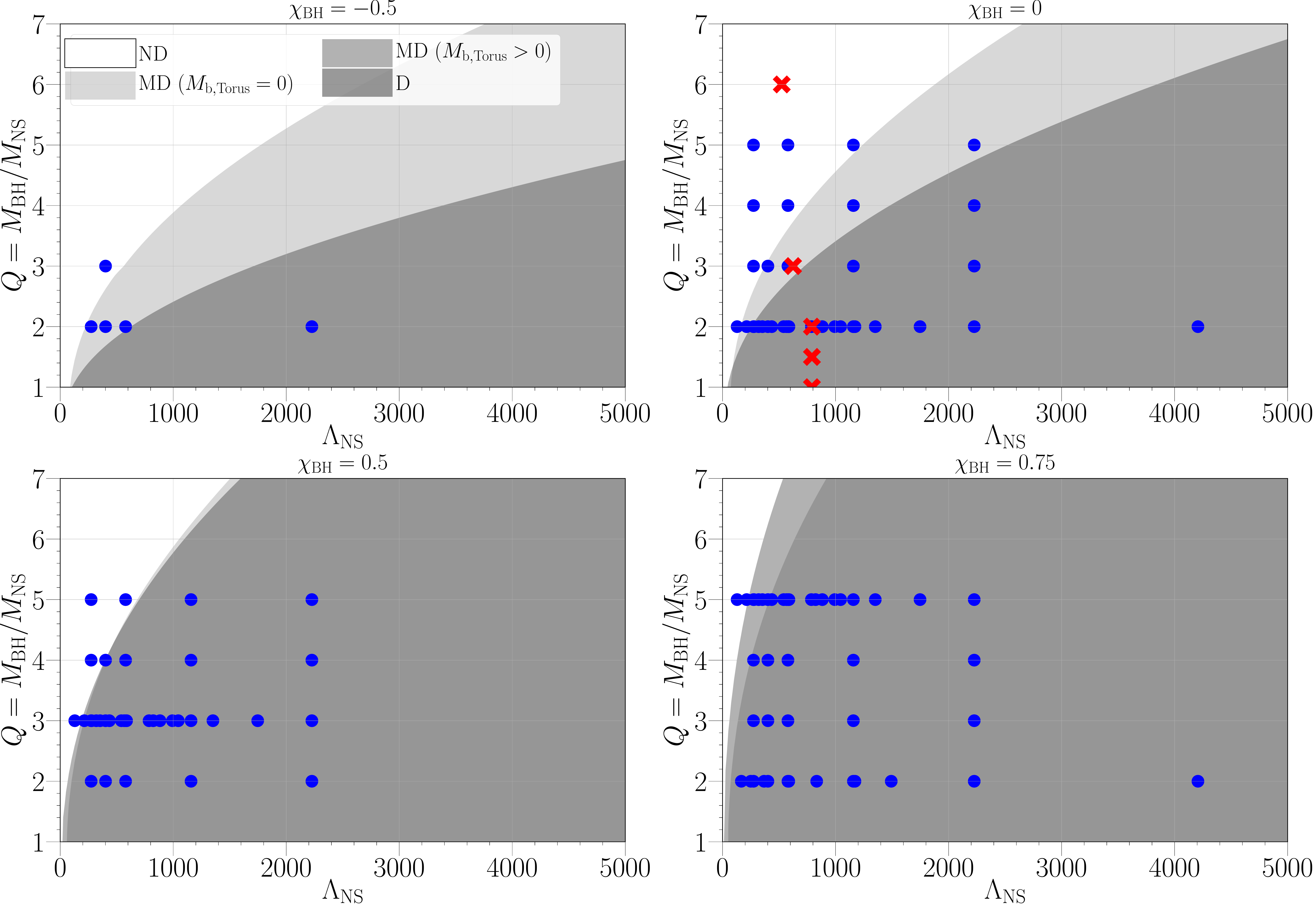}
    \caption{Representative parameter space region in the
      $Q-\Lambda_{\rm NS}$ plane for different values of the BH spin $\chi_{\rm BH}$. We show the regions for the
      different classes of NSBHs, using the model described in the
      main text: white regions represent non-disruptive mergers (ND),
      light gray regions represent mildly disruptive mergers without a
      torus remnant (MD, $M_{\rm b,torus}\neq 0$), the medium gray shade visible in the $\chi_{\rm
        BH}=0.75$ figure represents mildly disruptive mergers with a torus remnant (MD, $M_{\rm b,torus}= 0$), and the dark gray region marks
      disruptive mergers with a torus remnant (D). We also mark the
      parameter values of (some of) the NR simulations used. Dots
      (crosses) represent simulations produced by the {\tt SACRA}
      ({\tt SpEC}) code. Not shown are 12 {\tt SACRA} waveforms with
      $\chi_{\rm BH}=0.25$, two {\tt SpEC} waveforms with NS spin different from zero, and
      eleven 
     {\tt SpEC} waveforms with $\chi_{\rm
        BH}=\{0.7,0.8,0.9\}$.}
    \label{fig:Boundaries}
\end{figure*}

We limit the waveform modeling to the dominant quadrupolar multipole, notably 
the modes $\ell = 2, m = \pm 2$ in the -2 spin-weighted spherical harmonic decomposition of the gravitational 
polarizations $h_{+,\times}$, and to aligned-spin NSBHs. In the frequency domain, we can write the waveform as
\begin{equation}
    h(f) = A(f) e^{i \phi(f)}.
\end{equation}
Henceforth, we focus on the dependence of the amplitude $A(f)$ and phase $\phi(f)$ on the 
intrinsic parameters of the binary, $\vec{\theta}=\{M_{\rm BH},M_{\rm NS},\chi_{\rm BH},\chi_{\rm NS},\Lambda_{\rm
  NS}\}$, where we indicate with $\chi_{\rm BH}$ and $\chi_{\rm NS}$ the (dimensionless) components
of the spin aligned with the orbital angular momentum, for the BH and NS, respectively.

To compute the GW phase $\phi(f)$, we use the point-mass baseline \bbh\ model, and
apply tidal corrections from the NRTidal framework, as in Ref.~\cite{Dietrich:2019kaq}. 
As shown in Ref.~\cite{Foucart:2018lhe} (and as we verify in Figs.~\ref{fig:SXS-TD} and~\ref{fig:SXS-TD-2}), 
applying NRTidal corrections gives a reasonable approximation of the phase, until the last few
cycles.

In order to model the amplitude $A(f)$, we start with the BNS model
\bns\ as a baseline. Since this model includes tapering beyond the BNS
merger frequency \cite{Dietrich:2018uni}, we first remove this
tapering. This is necessary since the tapering depends on the tidal parameters
of both objects, $\Lambda_1$ and $\Lambda_2$, and does not vanish as $\Lambda_1\rightarrow 0$. We
note that this means that the $\Lambda_1\rightarrow 0$ limit of \bns\
does not correctly describe the amplitude of a NSBH system.

\begin{figure}
\includegraphics[width=0.49\textwidth]{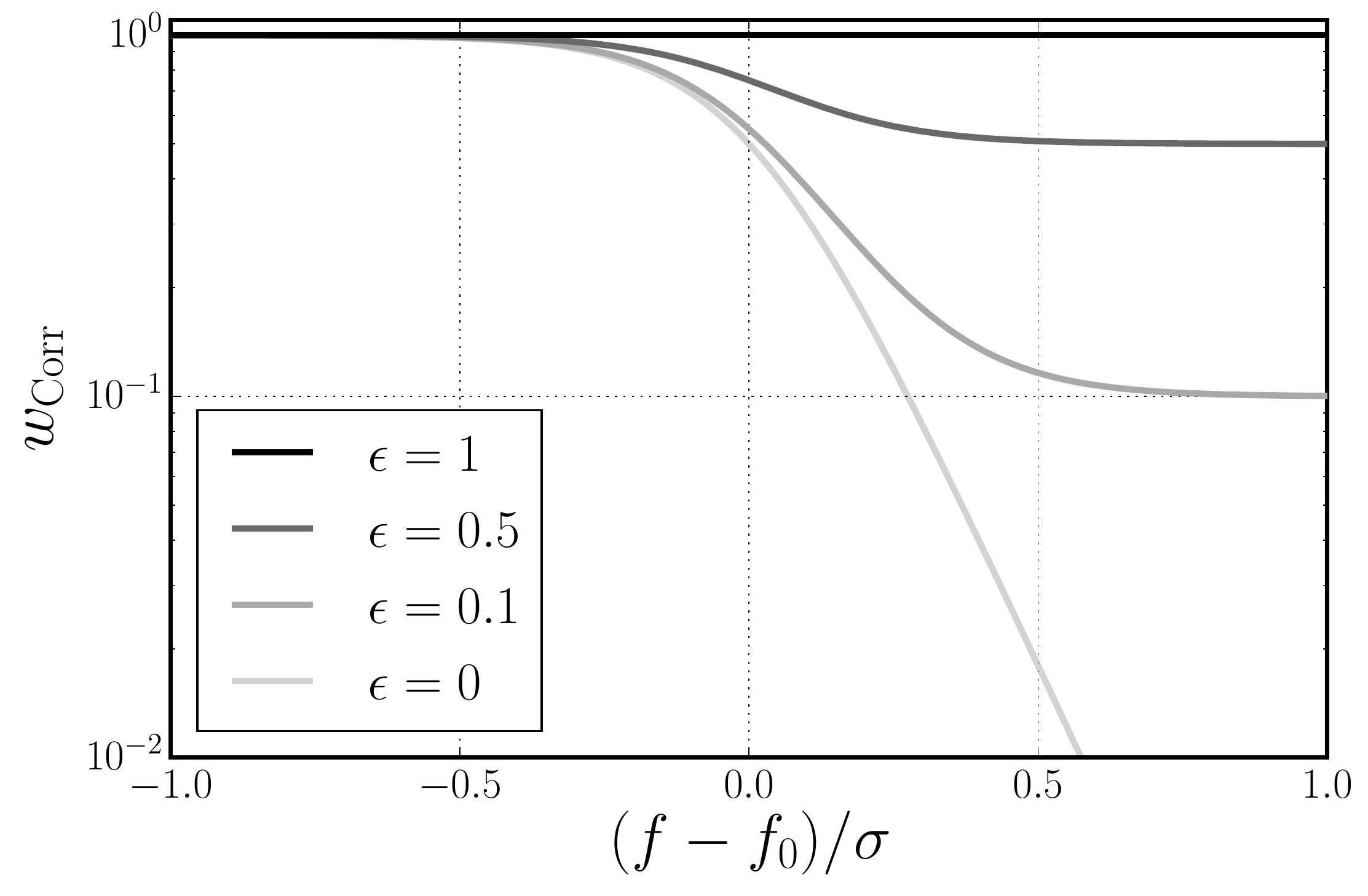}
\caption{We illustrate the behavior of the function $w_{\rm corr}(f)$ in Eq.~(\ref{amplitude}), 
which modifies the amplitude of the waveform with respect the one of a BNS and BBH.
The central frequency $f_0$ determines the frequency at
  which the amplitude is tapered; the width $\sigma$ determines the
  range of frequencies over which the tapering takes place; the
  parameter $\epsilon$ keeps the post-merger signal 
 at a suppressed level, if it is nonzero.}
\label{fig:wcorr}
\end{figure}

We then apply a correction to the amplitude that describes the tidal
disruption effects discussed in the previous section. More precisely,
we relate the amplitude of \nsbh, $A(f)$, to the amplitude of \bns\
with no tapering or tidal amplitude corrections applied, $A_{\rm NRT-notaper}(f)$, via
\begin{equation}
    A(f) = w_{\rm corr}(f) A_{\rm NRT-notaper}(f),
\label{amplitude}
\end{equation}
where the correction function $w_{\rm corr}$ is given by
\begin{equation}
    w_{\rm corr}(f) = w^{-}(f;f_0,\sigma) + \epsilon w^{+}(f;f_0,\sigma), 
\end{equation}
and $w^{\pm}(f;f_0,\sigma)$ are the hyperbolic-tangent window functions
\begin{equation}
    w^{\pm}(f;f_0,\sigma) = \frac{1}{2} \left[1\pm \tanh\left(\frac{4(f-f_0)}{\sigma}\right)\right].
\end{equation}
We illustrate the behavior of $w_{\rm corr}$ in Fig.~\ref{fig:wcorr}. When $\epsilon=0$, $w_{\rm corr}(f)$ cuts off
the amplitude before the end expected for a BBH system with the same masses and spins of the NSBH, and therefore describes tidal
disruption. When $\epsilon > 0$, the final part of the inspiral and the post-merger signal are still present, but are
suppressed relative to the BBH case. The parameters
$f_0,\sigma,\epsilon$, which determine the precise nature of these
corrections, are determined by comparing with NR simulations.

Following Ref.~\cite{Pannarale:2015jka}, we classify the waveforms
into four cases: non-disruptive, disruptive, mildly disruptive without
torus remnant, and mildly disruptive with torus remnant, depending on
the intrinsic parameters of the system. To determine the three parameters
\{$f_0,\sigma_0,\epsilon$\} in Eq.~(\ref{amplitude}), we adapt the amplitude model of
Ref.~\cite{Pannarale:2015jka}, which was developed for a different BBH
baseline, to the \bbh\ model. We then calibrate the parameters of
this model, using the method described in Sec.~\ref{sec:fits}.

\textbf{Non-disruptive mergers: $f_{\rm RD} < f_{\rm tide},
  M_{\rm b,torus}=0$.} When the tidal frequency is larger than the
ringdown frequency of the final BH, the NS reaches the ISCO before crossing $r_{\rm
  tide}$. In this case the NS remains intact as it plunges, but with a
slightly suppressed amplitude of the ringdown. The waveform is very
similar to a BBH. To describe this, we use $f_0=f_{\rm ND},
\sigma_0=\sigma_{\rm ND}, \epsilon=\epsilon_{\rm ND}$, where ND
stands for non-disruptive, and
\begin{equation}
\label{eq:wND}
w_{\rm ND}(f) = w^-(f;f_{\rm ND},\sigma_{\rm ND}) + \epsilon_{\rm ND}  w^+ (f;f_{\rm ND},\sigma_{\rm ND}).
\end{equation}
The explicit expressions relating $f_{\rm ND}, \sigma_{\rm ND}$, and $\epsilon_{\rm ND}$ to the intrinsic parameters of the binary are
given in Appendix~\ref{app:w_definition}.

\textbf{Disruptive mergers: $f_{\rm RD} > f_{\rm tide},
  M_{\rm b,torus}>0$.}  In this case, tidal disruption occurs and a
remnant torus of matter forms. For such systems, the typical merger
and ringdown stages present for BBHs are exponentially suppressed. To model
this case, we set $\epsilon=0$, so that the waveform decays above a
frequency $f_{\rm D}$ with width $\sigma_{\rm D}$. This leads to the expression
\begin{equation}
\label{eq:wD}
w_{\rm D}(f) = w^- (f;f_{\rm D},\sigma_{\rm D}).
\end{equation}
The precise definition is given in Appendix~\ref{app:w_definition}.

\textbf{Mildly disruptive mergers with no torus remnant:
  $f_{\rm RD} > f_{\rm tide}, M_{\rm b,torus}=0$.}  In this case, the
NS undergoes mass shedding, but no torus forms around the remnant
BH. We combine the information from the non-disruptive and disruptive cases to determine the
cutoff frequency and the width of the tapering. We set
$f_0=(1-Q^{-1})f_{\rm ND}+Q^{-1} f_{\rm tide},
\sigma_0=(\sigma_{\rm D}+\sigma_{\rm ND})/2, \epsilon=0$.
    
\textbf{Mildly disruptive mergers with torus remnant: $f_{\rm
    RD} < f_{\rm tide}, M_{\rm b,torus}>0$.} In this scenario 
the tidal frequency is above the ringdown frequency, but there is a remnant disk of matter
around the BH. As discussed, for example, in Ref.~\cite{Shibata:2011jka},
this scenario occurs at large BH spins, and represents the case in which the NS is disrupted before 
crossing the ISCO, but the size of the tidally disrupted material in the vicinity of the 
BH is smaller than the BH surface area. Thus, in this case, although 
a remnant disk eventually forms, the matter does not distribute uniformly around the BH quickly enough to 
cancel coherently or suppress the BH oscillations. As a consequence, the ending part of the 
NSBH waveform contains a ringdown signal.
In this case, we again combine
information from Cases 1 and 2, and fix $f_0=f_{\rm D},
\sigma_0=\sigma_{\rm ND}, \epsilon=\epsilon_{\rm ND}$.
    
In Fig. \ref{fig:Boundaries}, we show the regions of these different parameter spaces, 
along with relevant NR simulations from the \verb+SACRA+ and \verb+SpEC+ codes.

\subsection{Fitting procedure}
\label{sec:fits}

The amplitude correction described in the previous section has 20 free
parameters, which we denote with the vector $\vec{\lambda}$. The
definition of these parameters is given in
Appendix~\ref{app:w_definition}. We fix the coefficients in
$\vec{\lambda}$ by requiring that the \nsbh\ waveforms agree, 
as much as possible, with the \verb+SpEC+ and \verb+SACRA+ waveforms described in Sec.~\ref{sec:NR-data}.

\begin{table}[ht]
\centering
\begin{tabular}{c||c|c|c}
   \hline
   \hline
 Simulation  &$ f_{\rm low}$ [Hz] & \nsbh & PhenomNSBH    \\ 
 \hline 
  \texttt{SXS:BHNS:0001}  &  169 &   $7.5 \times 10^{-3}$ & $8.8 \times 10^{-3}$ \\
  \texttt{SXS:BHNS:0002} &  315 & $6.0 \times 10^{-3}$  & $4.5 \times 10^{-3}$   \\
  \texttt{SXS:BHNS:0003}  &  407  & $5.5 \times 10^{-3}$ & $5.2 \times 10^{-3}$  \\
  \texttt{SXS:BHNS:0004} &  447   & $7.8 \times 10^{-3}$  & $1.9 \times 10^{-2}$    \\
  \texttt{SXS:BHNS:0006}  &   314  & $5.6\times 10^{-3}$ & $5.3 \times 10^{-3}$   \\
    \texttt{M12-7-S8-LS220}  &  351 & $6.9\times 10^{-3}$ & $7.8 \times 10^{-3}$   \\
   \texttt{M12-7-S9-LS220}    &   343 & $1.1\times 10^{-2}$ & $7.4 \times 10^{-3}$ \\
    \texttt{M12-10-S8-LS220}  &   279 & $8.4 \times 10^{-3}$  & $2.0 \times 10^{-2}$ \\
     \texttt{M12-10-S9-LS220}  &   271 & $1.1 \times 10^{-2}$  & $2.7 \times 10^{-2}$  \\
      \texttt{M14-7-S7-LS220}   &   431  & $1.6\times 10^{-2}$ & $9.3\times 10^{-3} $ \\
       \texttt{M14-7-S8-LS220}   &   397  & $1.1 \times 10^{-2}$ & $2.0 \times 10^{-2}$  \\
      \texttt{M14-7-S9-LS220}   &   426 & $1.7\times 10^{-2}$ & $1.4 \times 10^{-2}$  \\
      \texttt{M14-10-S8-LS220}   &   286 & $1.1 \times 10^{-2}$& $4.6 \times 10^{-2}$   \\
      \texttt{M14-10-S9-LS220}   &   297 & $1.1\times 10^{-2}$ & $4.3 \times 10^{-2}$   \\
 \hline    
 \texttt{SXS:BHNS:0005}   & 448 & $7.4\times 10^{-2}$ & $8.0 \times 10^{-2}$ \\
 \texttt{SXS:BHNS:0007}  &  315   & $7.3 \times 10^{-3}$ & $1.1 \times 10^{-2}$    \\    
   \texttt{Q3S9}   &   300 & $8.2  \times 10^{-3}$ & $4.9 \times 10^{-3}$   \\
\texttt{Q4S9}    &   238 & $1.0\times 10^{-2}$ & $ 6.5 \times 10^{-3}$  \\  
   \hline
\end{tabular}
\caption{We list the unfaithfulness between different waveform models and SXS NSBH simulations. To compute the unfaithulness we use the Advanced LIGO design sensitivity PSD. In the left columns we show the unfaithfulness between \nsbh\ and PhenomNSBH, and the NR simulations. We compute the mismatch integral starting at the lower frequency $f_{\rm low}$ listed; this corresponds to a time late enough in the waveform that effects of junk radiation and eccentricity are negligible. The simulations below the horizontal line indicate simulations with nonzero $\chi_{\rm NS}$. These simulations were not included in the calibration and are used for validation.} 
    \label{tab:SXS_comparisons}
\end{table}

\begin{table*}[ht]
\centering
\begin{tabular}{c||c|c||c|c||c|c}
   \hline
   \hline
  \multicolumn{1}{c}{} &  \multicolumn{2}{c}{Window} & \multicolumn{2}{c}{${\rm SEOBNR\_NSBH}$-Hybrid} &  \multicolumn{2}{c}{PhenomNSBH-Hybrid }  \\
  \hline
 Simulation  & $t_{\rm min}$ [s] & $t_{\rm max}$ [s]&  \nsbh &   PhenomNSBH & \nsbh & PhenomNSBH   \\ 
 \hline 
  \texttt{SXS:BHNS:0001}  & 0.01 & 0.025 &  $1.2 \times 10^{-3}$  & $1.2\times 10^{-2}$  & $ 2.4 \times 10^{-3}$ & $1.3 \times 10^{-2}$ \\
  \texttt{SXS:BHNS:0002}   & 0.01 & 0.025  &   $7.4 \times 10^{-4}$ & $1.3 \times 10^{-4}$ & $1.3 \times 10^{-3}$  &  $1.4 \times 10^{-4}$ \\
  \texttt{SXS:BHNS:0003} & 0.005 & 0.018  & $2.1 \times 10^{-4}$  & $1.6 \times 10^{-4}$ & $2.7\times 10^{-3}$ & $1.3\times 10^{-3}$ \\
  \texttt{SXS:BHNS:0004}  & 0.008 & 0.02 &  $1.6 \times 10^{-4}$ & $5.7 \times 10^{-4}$ & $9.1 \times 10^{-4}$ & $2.7 \times 10^{-4}$   \\
  \texttt{SXS:BHNS:0006}  & 0.01 & 0.025   &  $1.8 \times 10^{-3}$ & $1.9 \times 10^{-3}$ & $4.6\times 10^{-3}$ & $3.9 \times 10^{-4}$  \\
    \texttt{M12-7-S8-LS220}   & 0.008   &  0.026 & $2.0\times 10^{-3}$ & $1.8 \times 10^{-2}$ & $1.8 \times 10^{-2}$ & $4.2 \times 10^{-3}$  \\
   \texttt{M12-7-S9-LS220}  & 0.01 & 0.03   &  $1.4 \times 10^{-3}$ & $5.0 \times 10^{-3}$ & $1.8 \times 10^{-2}$ & $3.2 \times 10^{-3}$ \\
    \texttt{M12-10-S8-LS220}    & 0.01 & 0.04 &  $7.1 \times 10^{-3}$ & $1.9 \times 10^{-2}$ & $6.2 \times 10^{-2}$ & $1.4 \times 10^{-2}$ \\
     \texttt{M12-10-S9-LS220}   & 0.01   & 0.04 &  $1.1 \times 10^{-2}$  & $2.9 \times 10^{-2} $ & $7.2 \times 10^{-2}$ & $2.5 \times 10^{-2}$  \\
      \texttt{M14-7-S7-LS220} & 0.0075 & 0.02  &  $9.1\times 10^{-4}$ & $5.3\times 10^{-3}$ & $5.5 \times 10^{-3}$ & $2.9\times 10^{-4}$  \\
      \texttt{M14-7-S8-LS220} & 0.006 & 0.018  & $3.7\times 10^{-4}$ & $9.1\times 10^{-3}$ & $2.6 \times 10^{-3}$ & $2.2 \times 10^{-3}$ \\
      \texttt{M14-7-S9-LS220} & 0.0075 & 0.02 &   $1.6\times 10^{-3}$ & $3.7 \times 10^{-3}$ & $7.1 \times 10^{-3}$ & $3.3 \times 10^{-4}$  \\
      \texttt{M14-10-S8-LS220} & 0.005  & 0.03  &   $1.4 \times 10^{-3}$ & $2.4 \times 10^{-2}$ & $3.6 \times 10^{-2}$ & $4.3 \times 10^{-3}$  \\
      \texttt{M14-10-S9-LS220}  & 0.007  & 0.03  &   $1.1 \times 10^{-2}$ & $6.5 \times 10^{-2}$ & $3.3 \times 10^{-2}$ & $4.3 \times 10^{-3}$  \\
 \hline    
 \texttt{SXS:BHNS:0005}    & 0.005 & 0.015   & $3.4 \times 10^{-3}$ & $4.0 \times 10^{-2}$ & $4.7\times 10^{-2}$ & $3.6 \times 10^{-3}$ \\ 
 \texttt{SXS:BHNS:0007}  & 0.005 & 0.055 & $ 6.4 \times 10^{-4}$ & $1.9 \times 10^{-2}$ & $7.1 \times 10^{-3}$ & $1.2 \times 10^{-3}$  \\   
   \texttt{Q3S9}    & 0.01 & 0.025 & $5.5 \times 10^{-4}$ & $1.2 \times 10^{-2}$ & $1.5 \times 10^{-2}$ & $3.3 \times 10^{-4}$  \\
\texttt{Q4S9} & 0.01 & 0.025  &  $ 8.6  \times 10^{-4}$ & $2.4 \times 10^{-3}$ & $6.9 \times 10^{-3}$ & $2.4 \times 10^{-3}$  \\   
  \hline
\end{tabular}
\caption{Unfaithfulness between the PhenomNSBH and \nsbh\ waveform models, and NR-hybrids constructed with these models. The details of the hybrid construction are given in the main text. We give here the initial $t_{\rm min}$ and final $t_{\rm max}$ times of the window used for hybridization in seconds, relative to the beginning of the NR data.} 
    \label{tab:SXS_hybrid_comparisons}
\end{table*}

For a given NR waveform indexed by I, let us denote the Fourier-domain amplitude of
the dominant mode by $A^{\rm NR}_{\rm I}(f;\vec{\theta})$. Given the intrinsic parameters of the binary,
$\vec{\theta}$, and a set of fit parameters
$\vec{\lambda}$, we compute the following quantity $\Delta_{\rm I}(\vec{\lambda})^2$,
\begin{equation}
\Delta_{\rm I}(\vec{\lambda})^2 = \int_{f_{\rm min}}^{f_{\rm cut}} {\rm d} f \frac{\left [A(f;\vec{\theta},\vec{\lambda}) - A_{\rm I}^{\rm NR}(f;\vec{\theta})\right]^2}{\sigma_{\rm I}(f)^2},
\label{Delta}
\end{equation}
to estimate the difference between the
frequency-domain amplitude of the model $A(f)$ and of the NR
simulation $A_{\rm I}^{\rm NR}(f)$. We choose the lower bound of the integral
$f_{\rm min}$ to be the frequency at which $A^{\rm NR}_{\rm I}(f)$ falls to 90\% of its initial (lowest-frequency) value; this is a low enough frequency to ensure $w_{\rm corr}(f_{\rm min})\approx 1$ while avoiding
possible contamination from eccentricity in the initial data. 
For the upper frequency, we take a definition inspired by the cutoff frequency given in Ref.~\cite{Pannarale:2015jia}.
First, we define $f_{\rm max}$ to be the frequency at which $f^2 A(f)$ takes its maximum value. Then we define $f_{\rm cut}$ to be the frequency (larger than $f_{\rm max}$) which satisfies
\begin{equation}
    f_{\rm cut} A_{\rm I}^{\rm NR}(f_{\rm cut}) =0.1  f_{\rm max} A_{\rm I}^{\rm NR}(f_{\rm max}).
\end{equation}
This frequency is larger than the ringdown frequency for non-disruptive mergers.
For disruptive mergers, $f_{\rm cut}$ gives a characteristic frequency at which the frequency domain amplitude has
been suppressed.
For the error function in Eq.~(\ref{Delta}), we consider a constant relative error at 
each frequency given by $\sigma_{\rm I}(f)=k_{\rm I} A_{\rm I}(f)$. We
use $k_{\rm I}=1$ for the \verb+SACRA+ waveforms, and $k_{\rm I}=0.1$ for the
\verb+SpEC+ waveforms, to account for the difference in length and
accuracy in the waveforms. We then compute a global error, for a given
subset $\mathcal{S}$ of the NR waveforms, by summing $\Delta_{\rm I}^2$ over
all NR waveforms in $\mathcal{S}$
\begin{equation}
    \Delta^2 (\vec{\lambda})= \sum_{ \rm I\in \mathcal{S}} \Delta_{\rm I}^2 (\vec{\lambda}).
\end{equation}
We minimize $\Delta^2(\vec{\lambda})$ with respect to $\vec{\lambda}$
using the Nelder-Mead algorithm \cite{Nelder-Mead}. We first use the
parameter values from Ref.~\cite{Pannarale:2015jka} as an initial
guess, and minimize the error over the parameters of the
non-disruptive and disruptive window functions
separately. We then use the results of this fit as an initial guess
for a global fit, including all of the available waveforms in
$\mathcal{S}$. The final results of this global fit are used to define
the model, and the numerical values are given in Table~\ref{tab:w_parameters} 
in Appendix~\ref{app:w_definition}.

\begin{figure}[ht]
    \centering
    \includegraphics[width=0.49\textwidth]{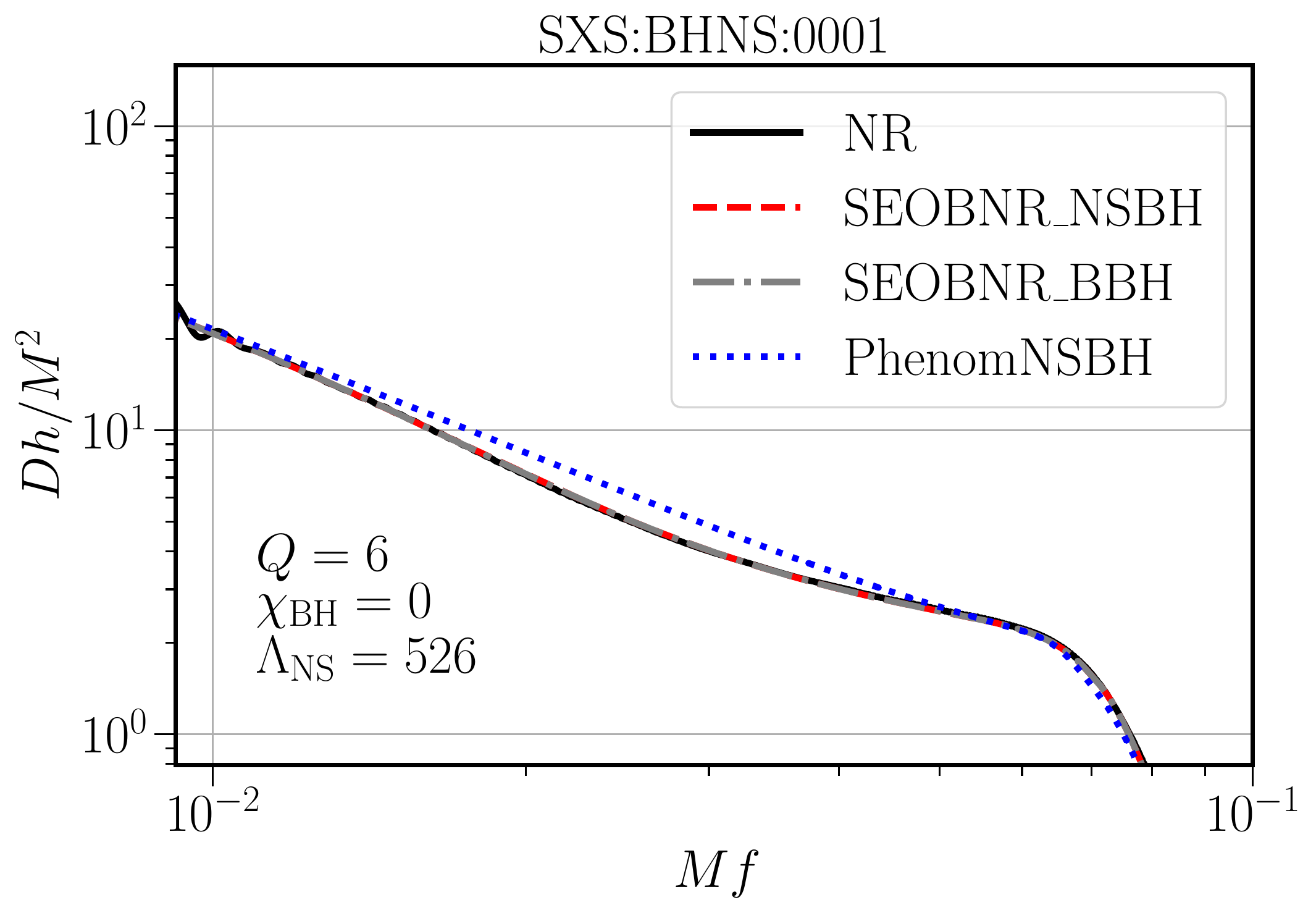}
   \includegraphics[width=0.49\textwidth]{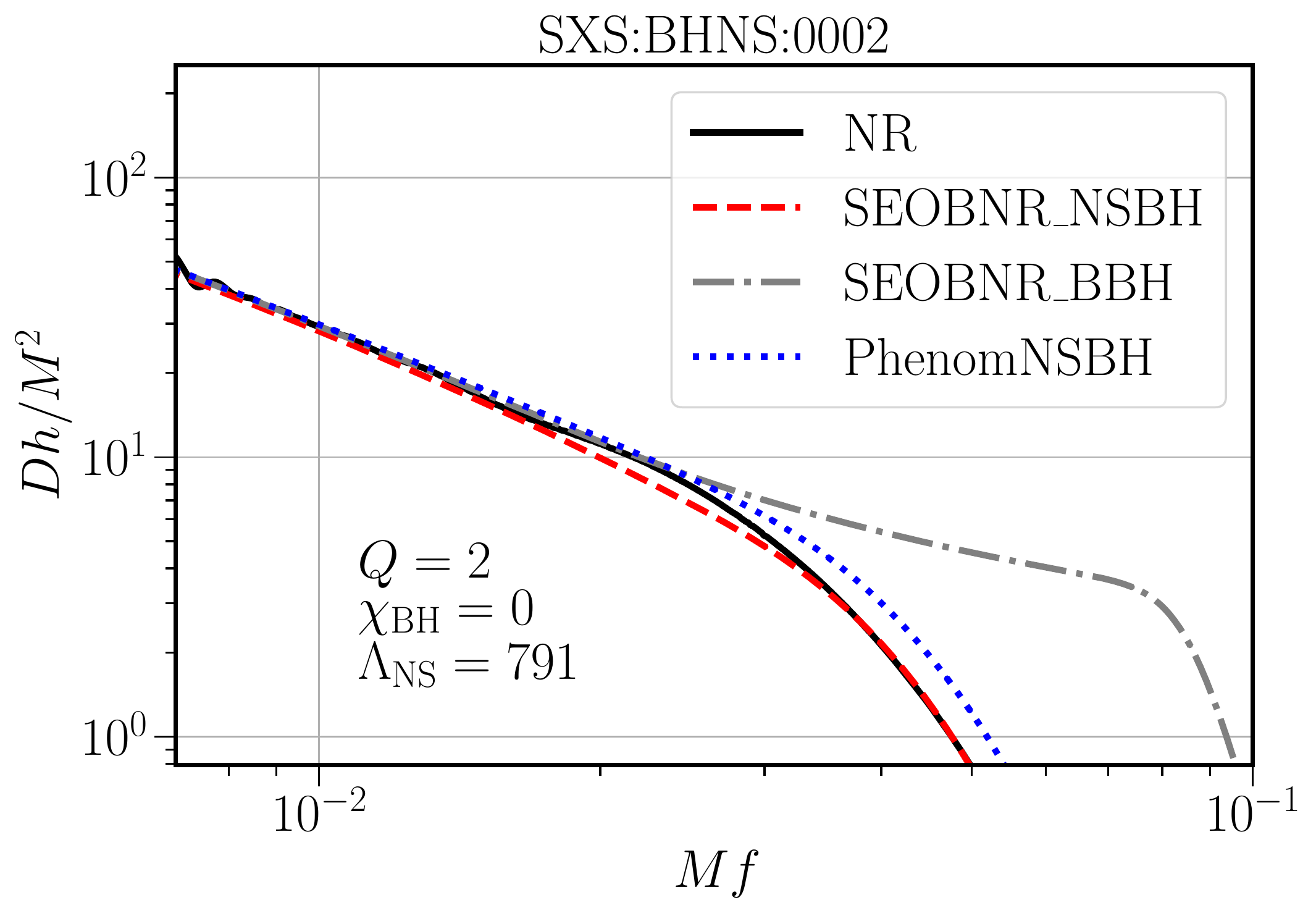}
    \caption{Frequency domain amplitude comparisons of SXS simulations, the NSBH waveform models \nsbh\ and PhenomNSBH, and the BBH model \bbh, that is used as a baseline for \nsbh. \nsbh\ is able to capture the effects of tidal disruption on the amplitude, while also reducing to BBH-like waveform for large mass ratios when tidal disruption does not occur.}
    \label{fig:SXS-FD}
\end{figure}

\begin{figure*}[ht]
    \centering
    \includegraphics[width=0.9\textwidth]{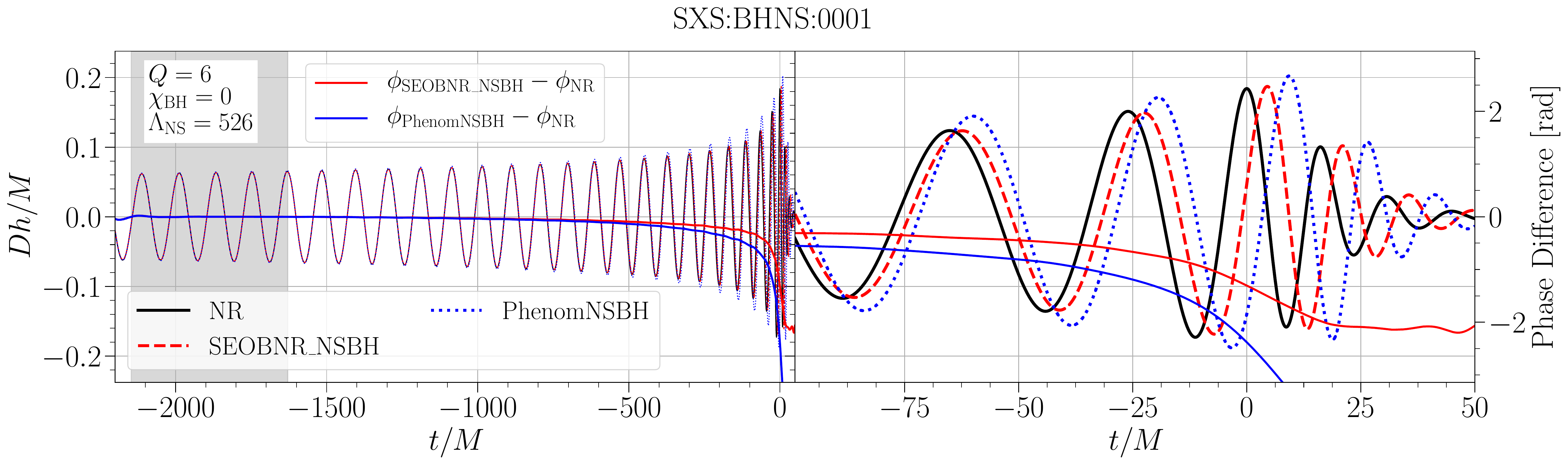}
    \includegraphics[width=0.9\textwidth]{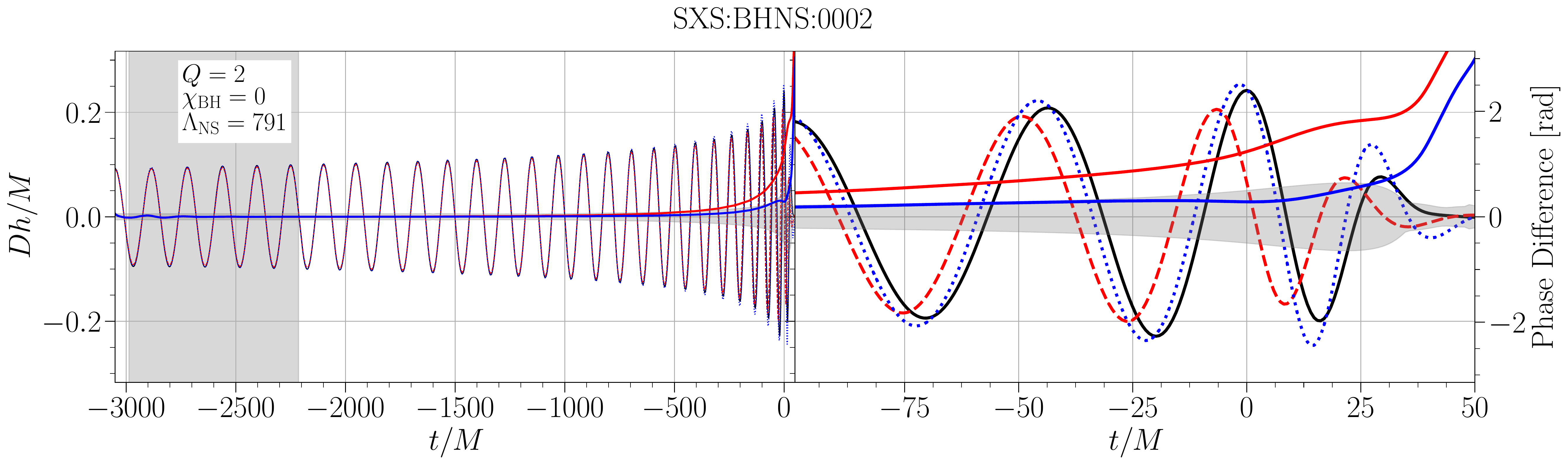}
    \caption{Time domain  comparisons of 2 NR simulations in the time domain, along with \nsbh\ and PhenomNSBH. We show  two of the publicly available SXS simulations with zero spin which were used for calibration of \nsbh: the non-disruptive merger \texttt{SXS:BHNS:0001}, with mass ratio 6, and the disruptive merger \texttt{SXS:BHNS:0002} with mass ratio 2. Also plotted is the phase difference for both NSBH models against the relevant NR simulation. The gray band shows the region used to align the model waveforms and NR. We also show the NR phase error for \texttt{SXS:BHNS:0002} as a horizontal gray band; for \texttt{SXS:BHNS:0001}, which is an older waveform, the NR error is not available. The NR waveform has been shifted in time so the peak amplitude occurs at $t=0$, and that the phase is zero there. We see that, for both waveform families, the agreement with NR is very good at the beginning of the NR waveform, but there is dephasing toward the end.}
    \label{fig:SXS-TD}
\end{figure*}

\begin{figure*}[ht]
    \centering
    \includegraphics[width=0.9\textwidth]{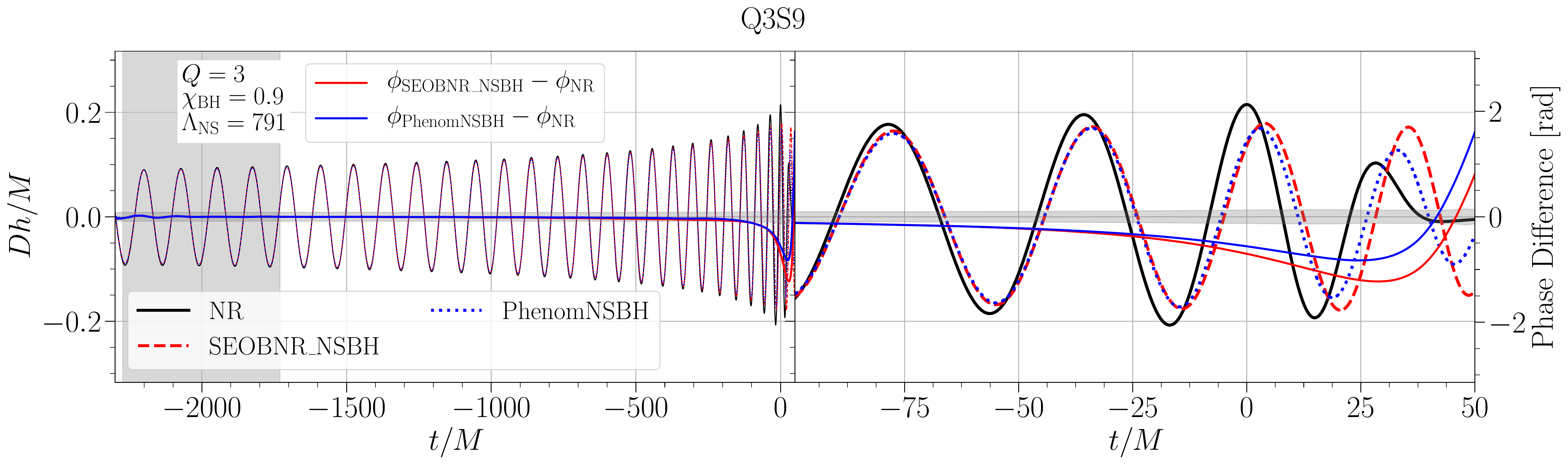}
     \includegraphics[width=0.9\textwidth]{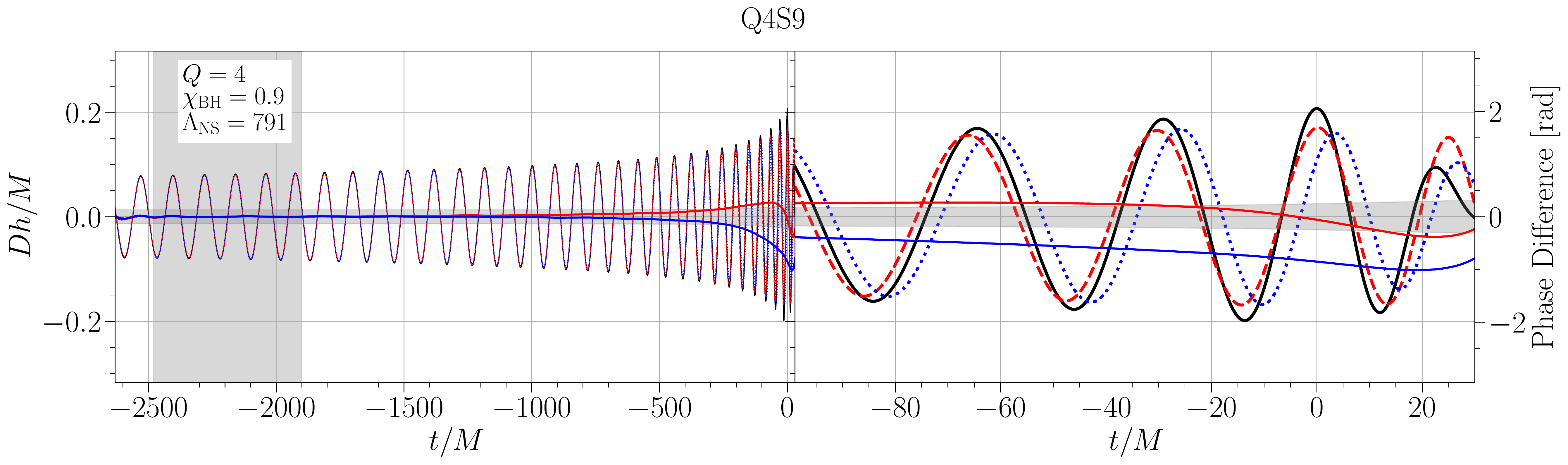}
    \caption{Time domain  comparisons of highly accurate waveforms of disruptive mergers, as well as the models \nsbh\ and PhenomNSBH.  We show  \texttt{Q3S9}, with mass ratio 3, and \texttt{Q4S9}, with mass ratio 4. Both configurations have a dimensionless BH spin magnitude of 0.9. These waveforms were not used to calibrate \nsbh. We see excellent agreement across a large number of cycles. The time and phase of the NR waveform have been fixed as in Fig.~\ref{fig:SXS-TD}.}
    \label{fig:SXS-TD-2}
\end{figure*}

We now turn to a quantitative assessment of the model's performance by comparing against NR simulations. We additionally compare with the recently developed PhenomNSBH model of Ref.~\cite{Thompson:2020nei}, in order to understand the performance of the two approximants relative to NR and to each other.
To this end, we employ the faithfulness function given in Ref.~\cite{Damour:1997ub}, 
which is commonly used in LIGO and Virgo data analysis to assess the agreement 
of two waveforms, e.g., the template $\tau$ and the signal $s$. Let us first introduce 
the inner product between two waveforms $a$ and $b$~\cite{Sathyaprakash:1991mt,Finn:1992xs}
\begin{equation}
    \langle a | b \rangle \equiv 4 {\rm Re} \int {\rm d} f \frac{a^*(f)\,b(f)}{S_n(f)}, 
\end{equation}
where a star denotes the complex conjugate 
and $S_n(f)$ is the one-sided, power spectral density (PSD) of the detector
noise. Here, we use the Advanced LIGO design sensitivity PSD as given in Ref.~\cite{LIGO_Design}. 
We compute the faithfulness $\mathcal{F}$ by maximizing the 
normalized inner product (or overlap) over the coalescence time $t_c$, the initial phase ${\phi_{0}}_\tau$ of the 
template $\tau$, and setting the phase of the signal ${\phi_{0}}_s$ to zero at merger, 
while fixing the same parameters $\vec{\theta}$ for the template and the signal, that is
\begin{equation}
    \mathcal{F} \equiv {\rm max}_{t_{c},{\phi_{0}}_\tau} \left [\frac{ \langle \tau | s \rangle}{\sqrt{\langle \tau | \tau   \rangle  
\langle s| s\rangle}} \right ]_{\left \{{\vec{\theta}}_\tau = {\vec{\theta}}_s,{\phi_{0}}_s=0 \right \}}.
\end{equation}
We find it convenient to discuss results also in terms of the unfaithfulness, that is 
$\bar{\mathcal{F}}=1-\mathcal{F}$. Henceforth, we 
consider the NR waveform as the signal, and the \nsbh\ or PhenomNSBH as the template. 
In Table~\ref{tab:SXS_comparisons} we list the unfaithfulness obtained against all the SXS NSBH waveforms 
at our disposal, for both \nsbh\ and PhenomNSBH. We also specify the lower frequency $f_{\rm low}$ used to compute the match. 
We see that both models have broadly similar performance.

Since the NR waveforms do not cover the entire bandwidth of the detector, 
we compute the faithfulness also between both NSBH waveform models, and NR hybrids.
We construct hybrids with both \nsbh\ and PhenomNSBH, and compare both waveform models
to the two hybrids. The four comparisons have two distinct purposes. 
First, the low frequency part of \nsbh\ and the \nsbh\ hybrid, and the PhenomNSBH and PhenomNSBH hybrid, are identical  up
to a shift in the time and phase of the waveform, so that the unfaithfulness of \nsbh\ with an \nsbh\ hybrid quantifies
the error of the waveform model failing to capture the NR; 
the same is true of the unfaithfulness between PhenomNSBH and a PhenomNSBH hybrid. 
Second, comparing \nsbh\ with a PhenomNSBH hybrid, and vice versa, includes the error from 
the NR part of the waveform, and additionally the error of waveform modeling uncertainty.
We show the results in Table~\ref{tab:SXS_hybrid_comparisons}. We note that the choice of hybrid affects the unfaithfulness: the unfaithfulness of SEOBNR\_NSBH with PhenomNSBH hybrids tends to be larger than the unfaithfulness of PhenomNSBH with SEOBNR\_NSBH hybrids.

To construct the hybrids, we follow the hybridization procedure given in Refs.~\cite{Dietrich:2018uni,Dietrich:2019kaq}.
We first align the waveforms by adjusting the time and phase of \nsbh\ to maximize the
overlap with the NR waveform, then we apply a Hann window to smoothly
transition from the model to the NR waveform. We refer to the initial and final times of the alignment window  as $t_{\rm min}$ and $t_{\rm max}$. These are chosen for each waveform to produce good agreement in the early part of the waveform. We provide the windows used in Table~\ref{tab:SXS_hybrid_comparisons}.

In Fig.~\ref{fig:SXS-FD}, we compare the frequency-domain amplitude of the \nsbh\ model and PhenomNSBH against two 
publicly available non-spinning SXS waveforms which were used to calibrate \nsbh. For context, we additionally show the BBH baseline model, 
\bbh. For \texttt{SXS:BHNS:0001}, which is a non-disruptive merger, the amplitudes of the NR data, \bbh, \nsbh,\ and PhenomNSBH,  agree well. 
For the disruptive merger \texttt{SXS:BHNS:0002}, \nsbh\ and PhenomNSBH capture
the tapering of the amplitude due to tidal disruption. In Fig.~\ref{fig:SXS-TD}, we compare 
 \nsbh\ and PhenomNSBH to the same two NR simulations in the time domain.
We include the NR error for those waveforms for which it is available, estimated using the methods described in Refs.~\cite{Foucart:2018lhe,Boyle:2019kee}.
In Fig~\ref{fig:SXS-TD-2}, we compare \nsbh\ and PhenomNSBH to the accurate spinning simulations
simulations \texttt{Q3S9} and \texttt{Q4S9}, which we use for validation.
We align the waveforms using the same procedure to construct the hybrids. 
We perform these comparisons using the $N=3$ extrapolation order.

\subsection{Regime of validity}
\label{sec:validity}

\begin{table}
\begin{tabular}{c|c|c}
  \hline
  \hline
Parameter & Calibration range & Suggested range of validity \\
\hline
$Q$ & [1,6] & [1,100] \\
$M_{\rm NS}$ & [1.2,1.4] $M_\odot$ & [1,3] $M_\odot$ \\
$\Lambda_{\rm NS}$ & [130,4200] & [0,5000] \\
$\chi_{\rm BH}$ & [-0.5,0.9]  & [-0.9,0.9] \\
  \hline
\end{tabular}
\caption{Range of intrinsic parameters for which the \nsbh\ waveform model was calibrated, and in which we suggest the model can be used. The calibration range gives the region of parameter space for which there are NR simulations. The suggested range of validity is the range we suggest for using the waveform, as explained in the main text. We note that a real EOS relates $\Lambda_{\rm NS}$ and $M_{\rm NS}$, and $\Lambda_{\rm NS}$ is a rapidly decreasing function of $M_{\rm NS}$. The largest value of $\Lambda_{\rm NS}$ for a 1.4 $M_\odot$ NS among the simulations we use is 791. See the main text for more detailed discussion.}
\label{tab:regime-of-validity}
\end{table}

In Table~\ref{tab:regime-of-validity}, we provide the parameter space region of the simulations used for calibration. We also give a suggested regime of validity for use of our \nsbh\  waveform model, which we justify as follows.
\begin{itemize}
\item \textbf{Mass ratio $Q$}. We take the lower limit for the
    mass ratio to be 1, given that in our fit we include NR
    simulations with these mass ratios. For large enough mass ratios,
    for any spin and $\Lambda_{\rm NS}$, the merger becomes non-disruptive and the model reduces
    to the \bbh\ waveform model. We have checked that there is always
    a range of parameter space at large mass ratios where this transition occurs,
    within the regime of validity of the model. Therefore we inherit the upper limit
    on $Q$ coming from \bbh, which is of 100.
\item \textbf{NS mass $M_{\rm NS}$.} Based on expectations of the
  maximum NS mass from the nuclear EOS, we restrict the NS mass to be less than $ 3 M_\odot$. 
We also suggest restricting the NS mass to be larger than $1 M_\odot$, 
which is consistent with the range that we choose for the  tidal 
parameter $\Lambda_{\rm NS}$.
\item \textbf{NS tidal-deformability $\Lambda_{\rm NS}$.} 
We have verified
that sensible waveforms are generated with $\Lambda_{\rm NS}$ varying from 0 up to 5000,
and on this basis suggest the waveform model can be used in this range. We have also performed a calibration and comparison against available NR simulations to verify the model accurately describes simulations with tidal disruption, as we have described. However the available NR simulations have a more limited range of $\Lambda_{\rm NS}$, depending on the NS mass and equation of state, as seen in Fig.~\ref{fig:Boundaries}. Thus we caution that tidal disruption effects are uncertain for large $\Lambda_{\rm NS}$, in particular $\Lambda_{\rm NS} \gtrsim1000$  for $1.4 M_\odot$ NSs. Even this restricted range includes the bound $\Lambda_{\rm NS}<800$ for a $1.4 M_\odot$ NS, obtained from measurements of GW170817  in Ref.~\cite{Abbott:2018exr}.

\item \textbf{BH spin $\chi_{\rm BH}$.} In the fit we 
  include simulations with positive spins as large as $0.9$, and
  negative spins as low as -0.5. Since negative spins tend to
  make the merger less disruptive (i.e., more BBH-like), in order to obtain
  a symmetric range we suggest $[-0.9,0.9]$ as a range for the spin.
\item \textbf{NS spin $\chi_{\rm NS}$.} While we do not include
  simulations with NS spin in the fit, from PN theory we
  expect that the main effect of the spin enters via the beta parameter derived in Ref.~\cite{Arun:2008kb}; here we use the formulation given in Eq A6 of Ref.~\cite{Ng:2018neg}
  \begin{equation}
  \beta = \frac{113-76\eta}{12} \chi_{\rm eff} + \frac{76\delta }{12} \chi_a
  \end{equation}
  where $\eta=Q/(Q+1)^2$ is the symmetric mass ratio, $\chi_a=(\chi_{\rm BH}-\chi_{\rm NS})/2$ is the antisymmetric combination of aligned spins, $\delta=(Q-1)/(Q+1)$ is an antisymmetric combination of the masses, and the effective aligned spin parameter
  $\chi_{\rm eff}$ is given by
\begin{equation}
\chi_{\rm eff} = \frac{M_{\rm BH} \chi_{\rm BH} + M_{\rm NS} \chi_{\rm NS}}{M_{\rm BH} + M_{\rm NS}}.
\end{equation}
Except for mass ratios close to 1 and small spins,
  $\beta$ is dominated by the BH spin. 
  We also see 
  reasonable agreement with simulations when the NS spin is nonzero,
as shown in Table~\ref{tab:SXS_comparisons}. We therefore recommend that the NS
  spin is bounded by the low-spin prior that has been used in the
  literature, e.g. Refs.~\cite{LIGOScientific:2018mvr,Abbott:2020uma}, $|\chi_{\rm
    NS}|<0.05$.
\end{itemize}
Through a thorough study, we have verified that the \nsbh\ waveforms 
look sensible in the region in which we suggest to use this model. 

\section{Applications}
\label{sec:Applications}

Having constructed the \nsbh\ waveform model and checked that it
agrees well with existing NR waveforms, we now apply the model to three
data-analysis problems. In particular, in Sec.~\ref{sec:fitting-factor}, we
compute the unfaithfulness of the \nsbh\ model against \bbh\ and \bns\ models in order to
obtain an estimate of the regions of parameter space where the
advanced-detector network may be able to distinguish different source
classes. In Sec.~\ref{sec:SXS-PE}, we perform a Bayesian parameter-estimation analysis in which 
we inject a synthetic NSBH signal (notably a disruptive NSBH merger) and infer the source's properties 
and parameter's biases when recovering it with the \nsbh\ model 
and the \bbh\ model. Finally, in Sec.~\ref{sec:GW170817}, we reanalyze the LIGO/Virgo 
event GW170817 under the hypothesis that it is a NSBH binary, instead of a BNS.

\subsection{Distinguishing different source classes}
\label{sec:fitting-factor}

When is it possible to determine whether a given binary system is a BBH, BNS,
or NSBH based on tidal effects in the gravitational waveform? We can address this
question with our waveform model by considering how similar a \nsbh\ waveform is
to a waveform from another source class. In this section we do not use an astrophysical prior on the masses of the objects to distinguish the source classes. Reference~\cite{Hannam:2013uu} considered the issue of distinguishing source classes, using measurements on the masses of the component objects, and an astrophyiscal prior on the masses of NSs and BHs, rather than measurements of the tidal parameter which we consider here. The conclusion of that work is that it will be difficult to distinguish different source classes with Advanced LIGO, with signals with SNRs in the range 10-20.

\begin{figure}[ht]
    \centering
    \includegraphics[width=0.49\textwidth]{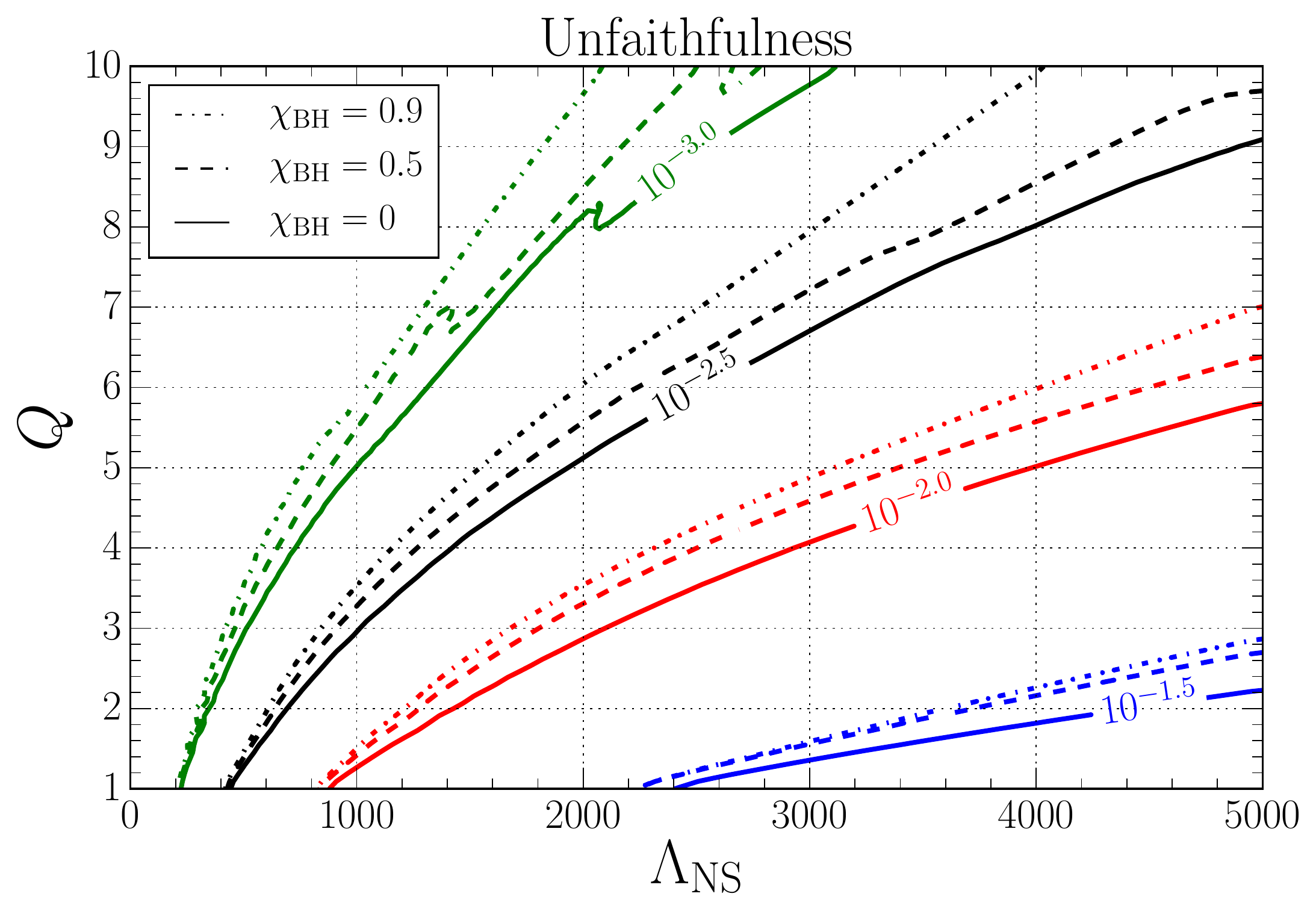}
    \caption{Contours with constant unfaithfulness in the $Q-\Lambda_{\rm NS}$ plane for varying BH spin $\chi_{\rm BH}$, when comparing \bbh\ and \nsbh\ assuming the Advanced LIGO design sensitivity PSD. As discussed in the main text, the unfaithfulness can be used to provide an estimate of the SNR at which data can be used to distinguish between two waveforms. Note that it is easier to distinguish BBH and NSBH systems for smaller $Q$ and larger $\chi_{\rm BH}$. }
    \label{fig:mismatches}
\end{figure}

First, we consider the case of distinguishing the hypotheses that a
given signal is a BBH or a NSBH. Suppose the signal is a NSBH with a
given set of parameters, $\vec{\theta}_{\rm NSBH}$. We compute the unfaithfulness 
between the \nsbh\ and \bbh\ models, with the same masses and spins. In the left
panel of Fig.,~\ref{fig:mismatches}, we show contours of the unfaithfulness 
in the $Q\mbox{--}\Lambda_{\rm NS}$ plane, for a $1.4 M_\odot$ NS and
$\chi_{\rm NS}=0$, while varying $\chi_{\rm BH}$ over the range $\{0,0.5,0.9\}$. To put the results in context,
following Refs.~\cite{Lindblom:2008cm}, we estimate that two waveforms
are distinguishable at the $1\mbox{--}\sigma$ level when the signal-to-noise
ratio (SNR) $\rho$ satisfies $\bar{\mathcal{F}}=K/2\rho^2$, where $K$ depends on the number of intrinsic parameters, $D$. 
Reference~\cite{Chatziioannou:2017tdw} provides an estimate of $K=D-1$, at which the $D$-dimensional posteriors do not overlap at the 1-$\sigma$  level. Then, an unfaithfulness 
of $\bar{\mathcal{F}}=10^{-3}$, corresponds to an SNR of $\rho\approx 45$. 
However, this criterion does not apply directly to marginalized posteriors. A more detailed discussion of the use of this criterion can be found in Ref.~\cite{Purrer:2019jcp}. In particular, the value of $K$ at which systematic errors become comparable to statistical ones, depend on what parameter is being considered, as well as extrinsic parameters such as the inclination.  They find that, when applying the criterion to marginalized posteriors, the estimate $K=D-1$ is conservative. Therefore, we emphasize that this criterion is sufficient, but it is not 
necessary, and also it does not say which parameters are biased 
and by how much. 

We also compute the unfaithfulness between the \nsbh\ and \bns\ models, with the same
masses, spins, and tidal parameters. We find that, for zero
spin, the unfaithfulness between NSBH and BNS are always less than
$10^{-3}$ when the NS mass is less than $3 M_\odot$. This suggests it
will be very difficult to distinguish NSBH and BNS systems on the
basis of tidal effects on the waveform alone. However, inference on the 
component masses provides additional useful information that can help 
distinguish different source classes.

As said above, computing the unfaithfulness does not allow us to quantify 
its impact on the inference of the parameters of the binary, and quantify 
possible biases. Therefore, in the next section, at least for one particular case, 
we perform a Bayesian parameter-estimation study and extract those biases, and 
compare with the distinguishability criterion of Refs.~\cite{Lindblom:2008cm,Chatziioannou:2017tdw}.

\subsection{Parameter-estimation case study}
\label{sec:SXS-PE}

In this section, because of computational costs, we perform a Bayesian 
parameter-estimation analysis for one specific NSBH system, 
and postpone to the future a more comprehensive analysis.

We first create a synthetic NSBH signal consisting of an NR hybrid built 
by stitching together the \nsbh\  waveform to the \texttt{SXS:BHNS:0006} waveform, with masses $M_{\rm BH}=2.1 M_\odot$, $M_{\rm NS}=1.4 M_\odot$, mass ratio $Q = 1.5$, and both spins equal to zero. We do not add a noise realization (i.e., we work in zero noise) which is equivalent to averaging over different noise
realizations, as shown in Ref.~\cite{Nissanke:2009kt}. We perform four injections, with SNRs of 25, 50, 75, and 100 in the advanced LIGO-Virgo network.
While the masses are not astrophysically motivated, this system is interesting to study because it is disruptive, and due to the mass ratio the tidal dephasing is enhanced. Further, \texttt{SXS:BHNS:0006} is the simulation with the largest number of cycles of the publicly available SXS waveforms.

We apply the Markov Chain Monte Carlo (MCMC) sampling algorithm
implemented in LALInference \cite{Veitch:2014wba} to these four
signals, and recover the signal with both the \bbh\ and \nsbh\
waveform models.  Due to limited computational resources, we run the
parameter estimation with a lower cutoff frequency of 30 Hz. We take the higher cutoff frequency to be 2048 Hz. We use a
uniform prior on the detector frame component masses. For \nsbh, we
impose a constraint that $M_{\rm NS}<3M_\odot$, $|\chi_{\rm
  NS}|<0.05$, and $|\chi_{\rm BH}|<0.9$ consistent with the range of
validity of the model. We take a prior on $\Lambda_{\rm NS}$ that is uniform
between 0 and 5000. For \bbh, we do not impose a constraint on the
maximum mass but do require that the spins of both objects were less
than 0.9. Since the two approximants make different assumptions about
the nature of the component objects, in describing the results of the
Bayesian analysis, we refer to the masses as $m_1$ and $m_2$ rather
than $M_{\rm BH}$ and $M_{\rm NS}$.

\begin{figure*}[ht]
    \centering
    \includegraphics[width=0.49\textwidth]{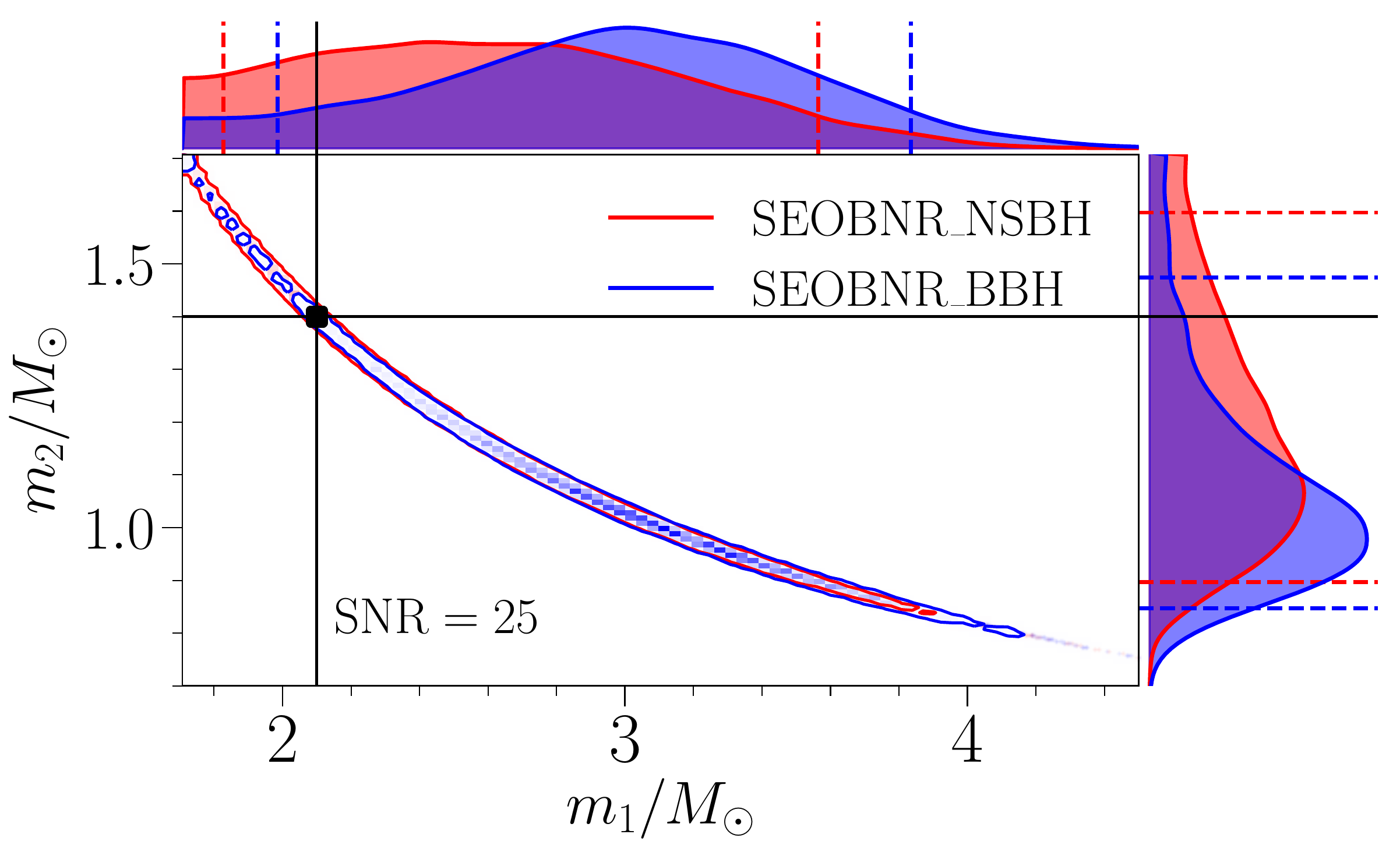}
    \includegraphics[width=0.49\textwidth]{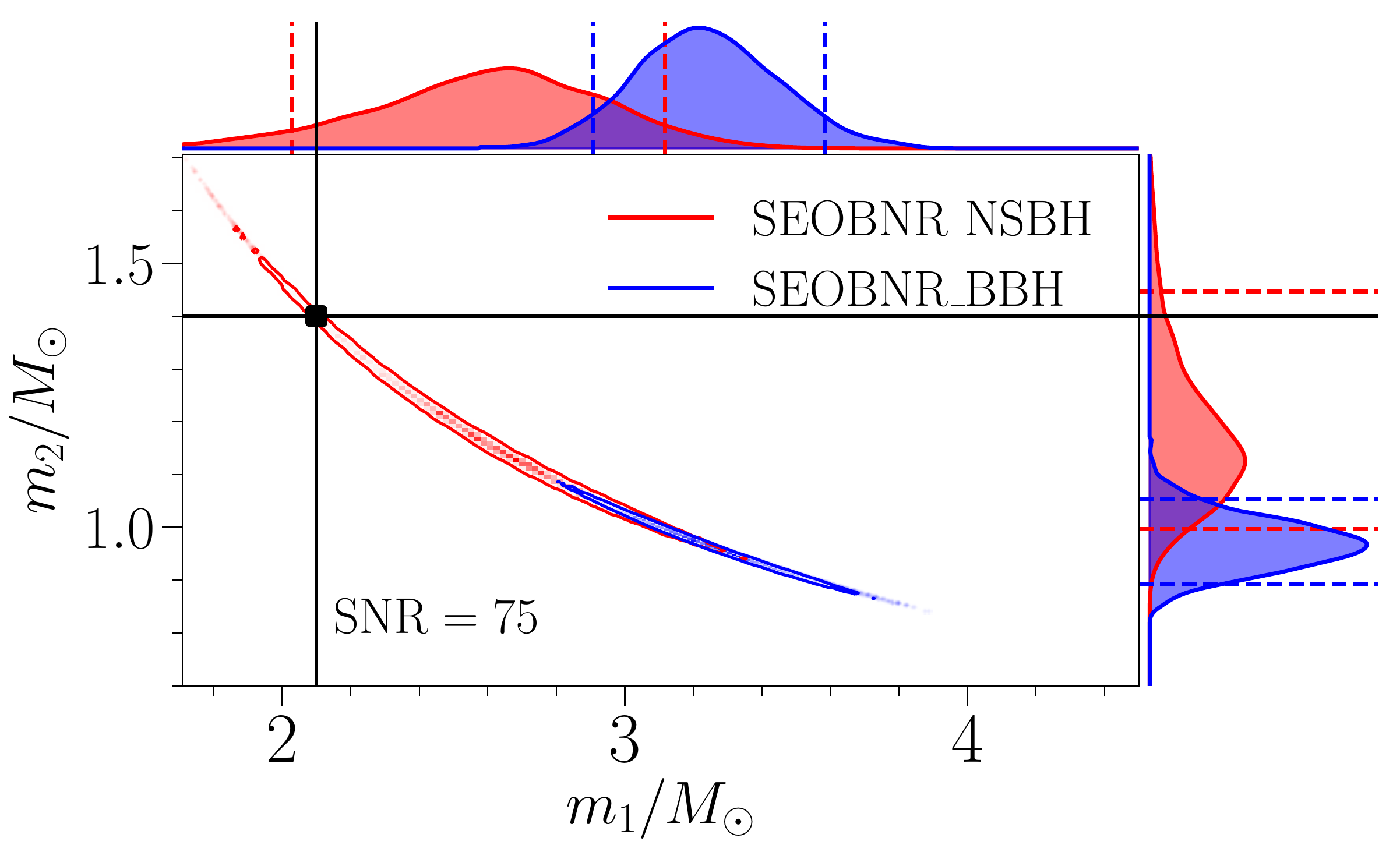}
    \includegraphics[width=0.49\textwidth]{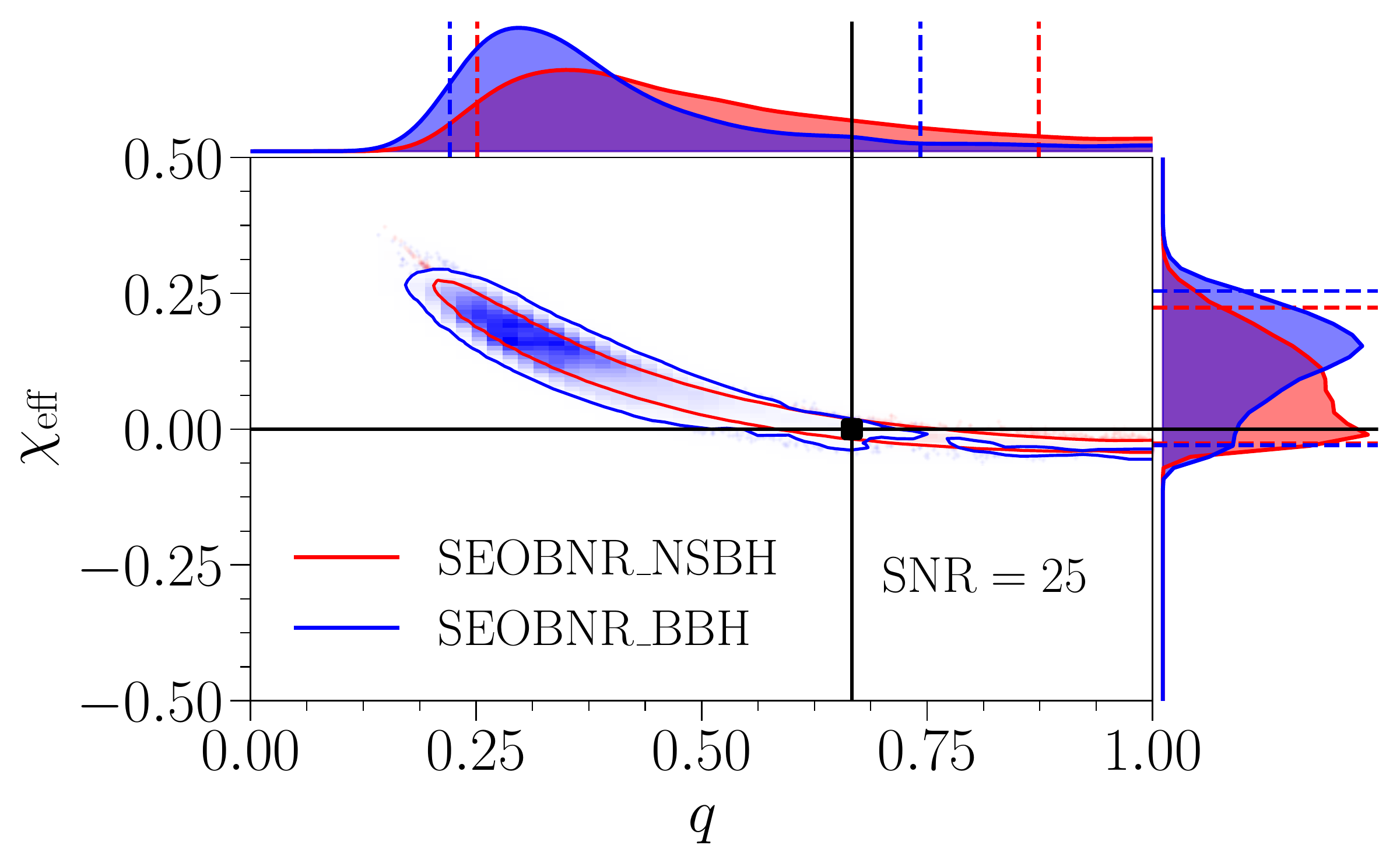}
    \includegraphics[width=0.49\textwidth]{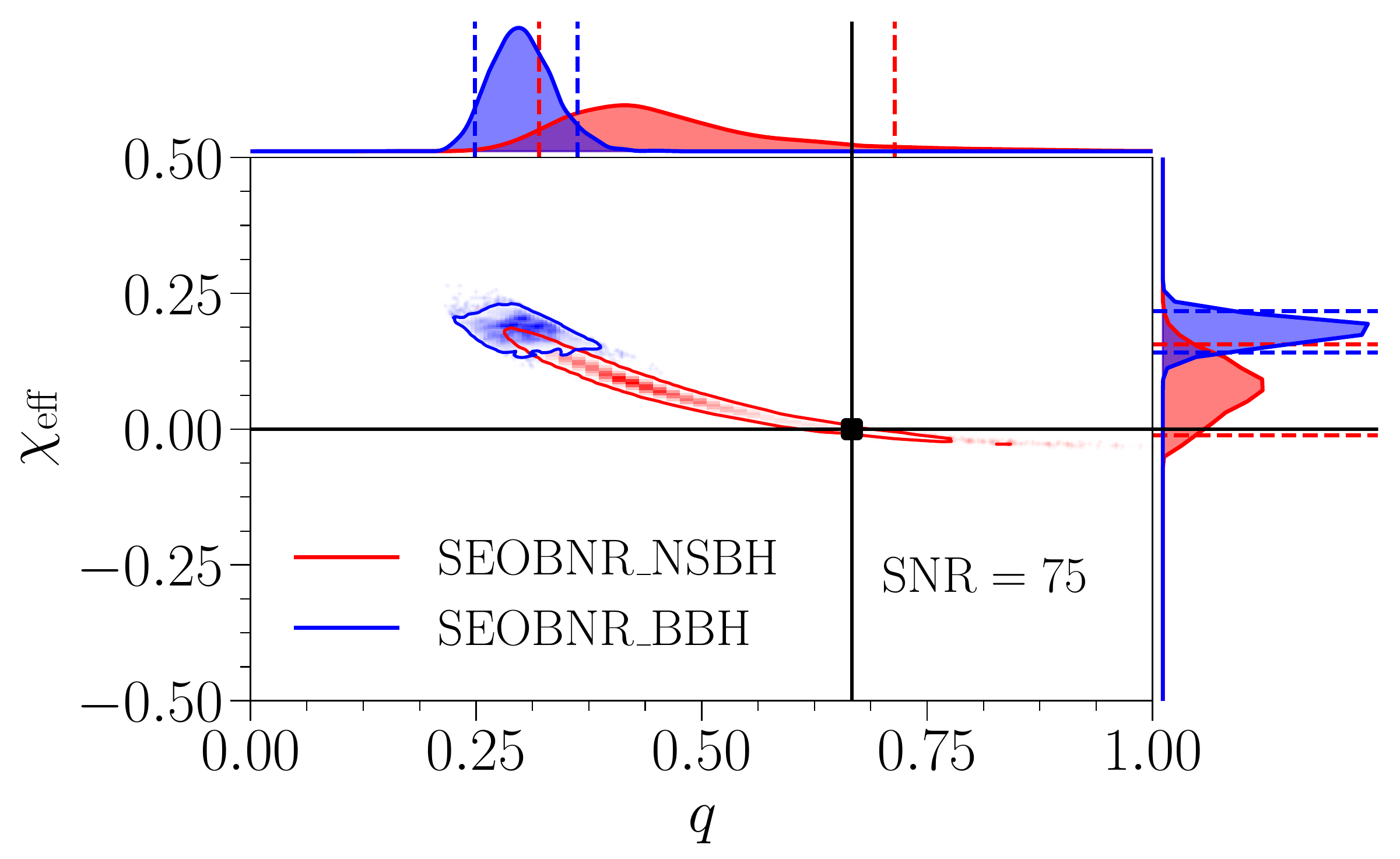}    
        \caption{ Illustrative parameter-estimation results for the
          \texttt{SXS:BHNS:0006} hybrid injections described in the main
          text. In the top two panels, we show posterior
          distributions and 90\% credible intervals for the component masses, with the posteriors derived using the
          \nsbh\ (\bbh) approximant in red (blue).  In the bottom panels,
          we show the $q-\chi_{\rm eff}$ plane. We show the
          injected value as a black dot. 
          For SNR=25, both \nsbh\ and \bbh\ recover the injected
          value within the 90\% credible interval. For larger SNRs, the recovery with
          \bbh\ is biased. The posterior with the BBH waveform is peaked around a larger total mass,
          than the injected one. 
          We show this
          explicitly for SNR=75. Additionally, we use this software injection test to explore how 
          the difference between
          \nsbh\ and the NR hybrid affects parameter estimation.
          We see that, at SNR=75, the injected values of the masses and $\chi_{\rm eff}$ are marginally consistent
          with the \nsbh\ posterior at the 90\% level; at larger SNRs we find the 90\% credible interval does not
          include the true value. } 
    \label{fig:Injection-1}
\end{figure*}

In Fig.~\ref{fig:Injection-1}, we show
posteriors in the $m_1-m_2$ plane, as well as the $q-\chi_{\rm eff}$ plane, for the SNR=25 and SNR=75 injections. For ease of comparison with other PE results by LIGO-Virgo analyses, we show the posterior in terms of the mass ratio $q\equiv Q^{-1}=m_2/m_1$. For the SNR=25 injection, we see the posteriors from the two waveforms agree very well and are consistent with the injected value within the 90\% credible interval. 
For larger SNRs, posteriors derived using the two waveforms are in tension, and at large enough SNR,
the injected value lies outside of the 90\% credible interval of the posterior for each model.
For the SNR=50 injection, and for larger SNRs, there is a bias in the masses and $\chi_{\rm eff}$ recovered using \bbh.
In particular, \bbh\ recovers a larger total mass.
The biases in the mass are due to the lack of tidal effects in \bbh. To quantify this, we have performed a run with two modified versions of \nsbh. The first modified model has a tidal phase correction but the same amplitude as \bbh. The second one has the tidal disruption correction to the amplitude but no tidal phase is applied. At SNR=25, both models recover the injected mass ratio inside  of the 90\% credible interval. At SNR=75, we find that the first modified model, like \nsbh, obtains the correct value within the 90\% interval. On the other hand,  the second model, with only the amplitude correction, does not. This is consistent with the fact that the tidal phase accumulates over many cycles, while the merger frequency is at relatively high frequencies outside of the most sensitive band of the detector.\footnote{We thank the anonymous referee for suggesting that the bias in this case can be attributable to the tidal phase.}
The masses and spins recovered by \nsbh\ are consistent at the 90\% level with the injected values for the SNR=50 case, 
but are only marginally consistent  for the SNR=75 injection, and for
SNR=100, the injected values of the masses and $\chi_{\rm eff}$ 
lie outside 90\% credible interval. This bias is 
due to differences with the NR-hybrid waveform.
We show the recovery of the SNR=75 injection, for which the \nsbh\ recovery is marginally consistent with the true parameters,
 in the right two panels of Fig.~\ref{fig:Injection-1}.

\begin{figure}[ht]
    \centering
    \includegraphics[width=0.49\textwidth]{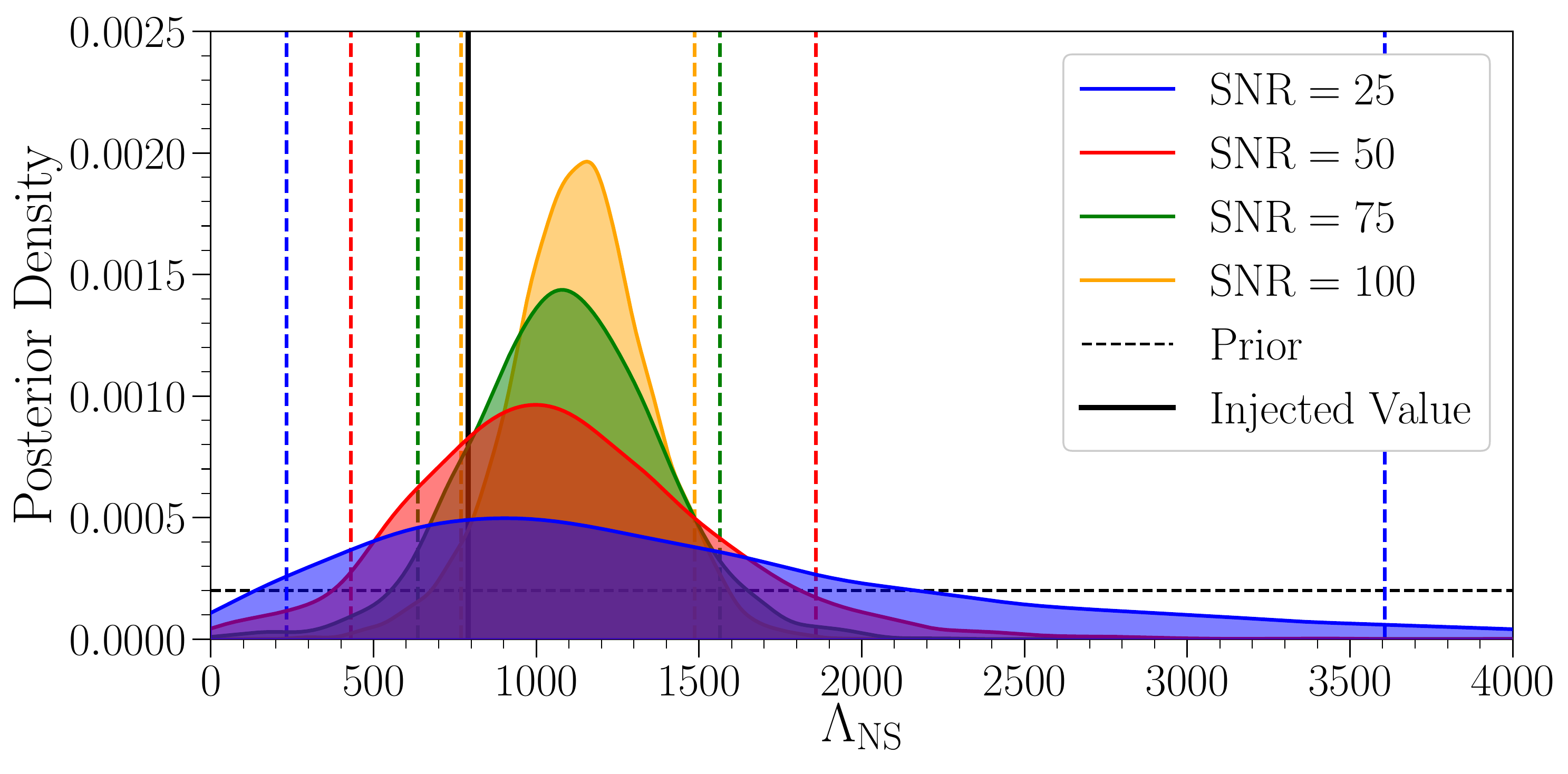}
        \caption{ Posteriors for the NS tidal
          deformability $\Lambda_{\rm NS}$ for the NSBH hybrid injections described in the main text, recovered with \nsbh,
          for different SNRs. The recovered values of $\Lambda_{\rm NS}$ are consistent with the true
          value from the hybrid at the 90\% level, however this is only marginally true for SNR=100.}
    \label{fig:Injection-2}
\end{figure}

In Fig.~\ref{fig:Injection-2}, we show recovery of the tidal parameter
$\Lambda_{\rm NS}$ obtained using \nsbh\ for the 4 different cases. 
In all four scenarios, the injected tidal parameter is consistent with the 90\% credible
interval of the $\Lambda_{\rm NS}$ posterior, although this is only marginally true for the
SNR=100 injection.
It is interesting to compare the difference between the recovered and injected values, with what is expected from
the indistinguishability criterion discussed in the previous section.
The unfaithfulness from 30 Hz between the NR hybrid used and \nsbh\ is $2 \times 10^{-3}$,
 using the advanced LIGO design sensitivity
PSD. From the indistinguishability criterion of Ref.~\cite{Chatziioannou:2017tdw} discussed in the previous section, 
we would expect to see deviations at the $1-\sigma$ level
between the posterior recovered with \nsbh\ 
and the injected value an SNR of 32.
A full Bayesian analysis reveals that this level of bias for the recovery of $\tilde{\Lambda}$ only arises at a larger value of the SNR. 
However as we have emphasized, the criterion 
strictly applies only to the full $D$ dimensional posterior, and not the marginalized posteriors we consider in this section. Additionally, the criterion is only sufficient, it does not specify which parameters are biased, depends on extrinsic parameters such as the inclination,
and it has been shown to be quite conservative when applied to the marginalized posteriors~\cite{Purrer:2019jcp}.

This case study illustrates the
importance of having accurate NSBH models that can account for tidal
disruption in order to derive correct conclusions about astrophysical parameters. However, we emphasize that these injections are only meant as an
example. Larger mass ratios may be less tidally disruptive and have
tidal effects on the phase suppressed. Conversely, systems with large BH spin will
tend to be more disruptive, which will enhance the differences between
the BBH and NSBH waveforms.

After this manuscipt was submitted, Ref.~\cite{Huang:2020pba} appeared as a preprint. This work provides a detailed parameter estimation study, recovering many different NSBH injections using \nsbh, PhenomNSBH, \bbh, \bns, and other waveform models. The injected waveforms are hybrids of \texttt{SXS:BHNS:0001}, \texttt{SXS:BHNS:0003}, and \texttt{SXS:BHNS:0004}, with the NR surrogate model \texttt{NRHybSur3dq8Tidal}, developed in Ref.~\cite{Barkett:2019tus}, at SNR=30 and SNR=70. Of the cases they study, \texttt{SXS:BHNS:0004}, with $M_{\rm NS}=M_{\rm BH}=1.4 M_\odot$, has the most similar parameters to the signal we consider in this work. The results that Ref.~\cite{Huang:2020pba}  obtains for \texttt{SXS:BHNS:0004} are broadly similar to the results we present here. When SNR=30, the injected masses are recovered within the 90\% credible intervals when recovering with \bbh, \bns,  \nsbh, and PhenomNSBH. When SNR=70, the component masses recovered using \bbh\ are biased towards larger values. On the other hand, the posteriors obtained with \bns, \nsbh\, and PhenomNSBH recover the component masses within the 90\% credible interval. The fact that \bns\ and \nsbh\ both recover the correct mass may indicate that the tidal phase is more important than the tidal disruption frequency for recovering the masses. For the tidal parameter, at SNR=70, the authors find that the recovery of \nsbh\, as well as \bns, is in tension at the 90\% level.  Interestingly, PhenomNSBH recovers the correct tidal parameter when SNR=70. This paper, like the current work, illustrates the need for accurate NSBH models as the detector network sensitivity improves.

\subsection{Inference of GW170817 as a NSBH}
\label{sec:GW170817}

As a final application, we reanalyze GW170817
\cite{TheLIGOScientific:2017qsa} under the hypothesis that it is a 
NSBH (see also Ref.~\cite{Hinderer:2018pei,Coughlin:2019kqf} for related
studies). Indeed, it is interesting to ask whether GW data
alone can be used to distinguish the hypotheses that this event is a
BNS or a NSBH. 

We run the Bayesian inference study with the  MCMC code implemented in LALInference, 
using publicly  available data of GW170817 from the GW open science center (GWOSC)
\cite{Vallisneri:2014vxa} (discussion of these data for O1 and O2 are
contained in Ref.~\cite{Abbott:2019ebz}). We run with both the \nsbh\ model, as well
as the \bns\ model in order to be able to do a fair comparison. 
As far as we know, this is the first time that the new version of the \bns\ model has been used to analyze GW170817. We
compare our results to those from the runs obtained in the GWTC-1
catalog \cite{LIGOScientific:2018mvr}, which used a former version of
the \bns\ model. 
We use the same priors as the GWTC-1 analysis \cite{LIGOScientific:2018mvr}, 
except where otherwise stated.
For \bns\ we assume a flat prior on $\Lambda_1$ and $\Lambda_2$,
while for \nsbh\ we assume a flat prior on $\Lambda_2$ and fix $\Lambda_1$ to zero. 
The posteriors contain support 
only in the interior of the prior domain for both waveform models with these priors. 
The prior on the component mass ranges from $0.5-7.7 M_\odot$, and therefore the 
prior does not require that both objects have masses below the maximum mass of a NS.

\begin{figure}
    \centering
    \includegraphics[width=0.49\textwidth]{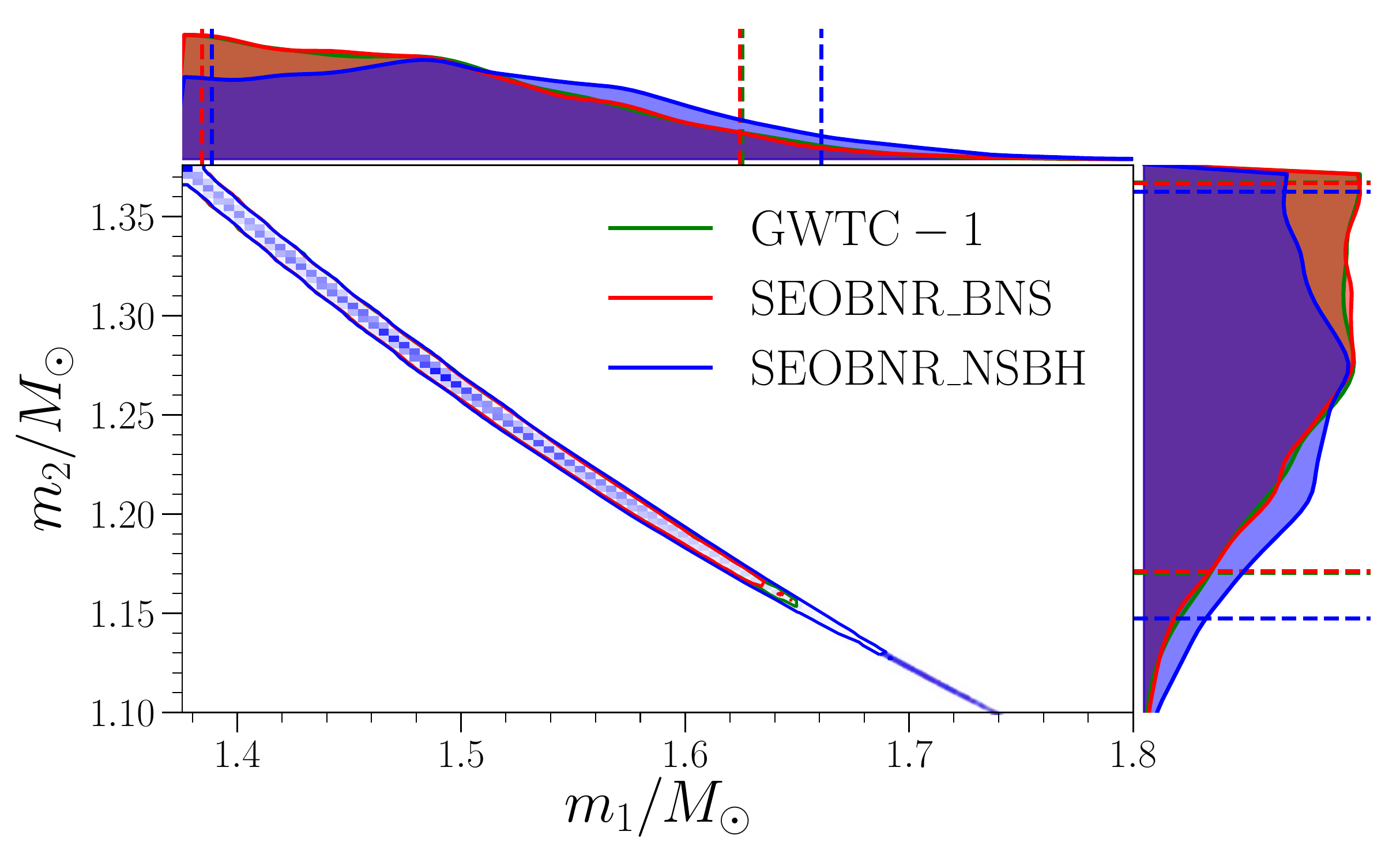}
    \includegraphics[width=0.49\textwidth]{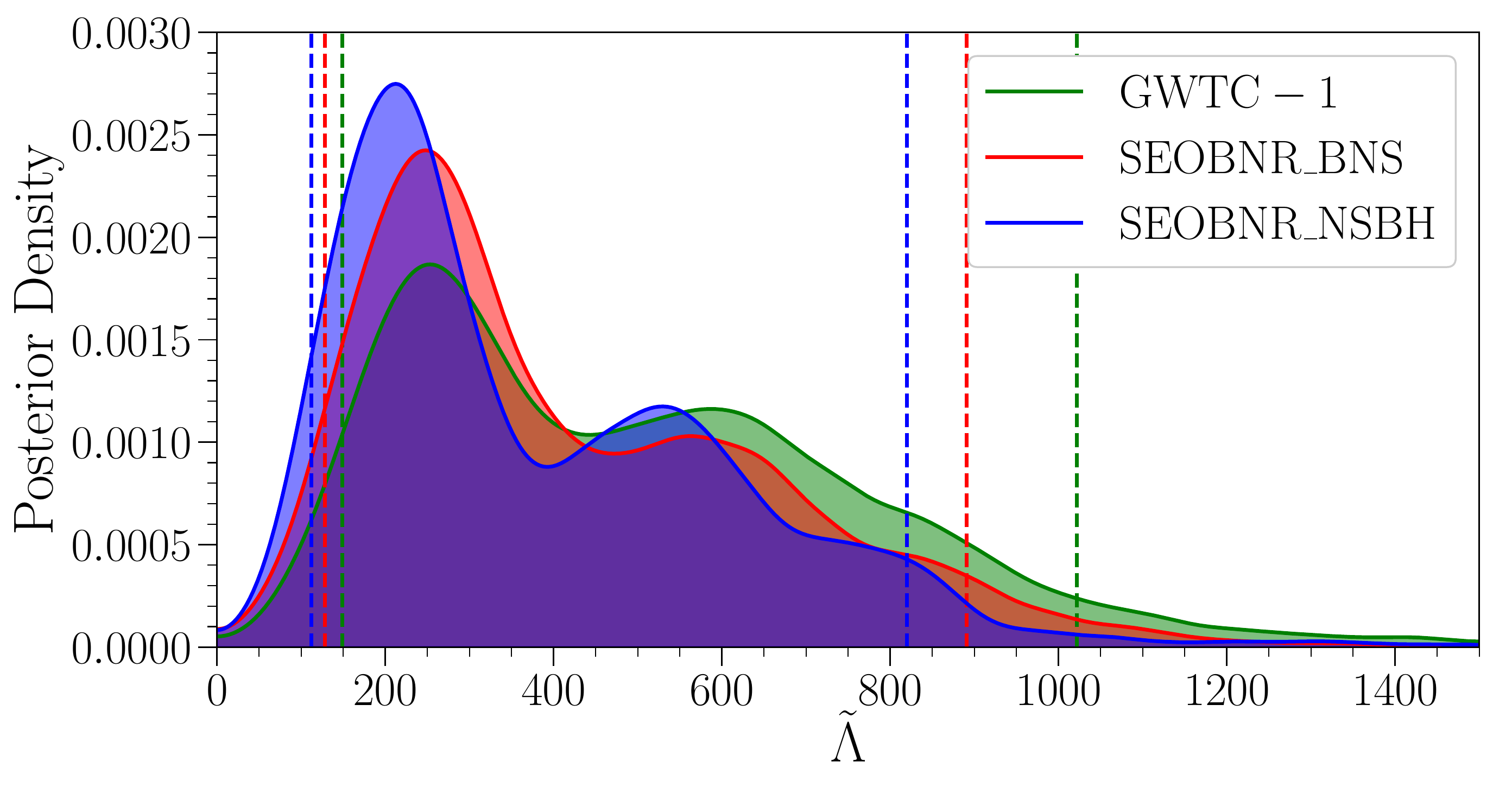}
    \caption{Reanalysis of GW170817. We show posteriors for the effective tidal deformability $\tilde{\Lambda}$ and the mass ratio $q\equiv m_2/m_1$, for three waveform models: in green we show the recovery with an older version of \bns\ which performed in GWTC-1. In red we show a recovery with the current version of \bns, which has been recalibrated in Ref.~\cite{Dietrich:2019kaq}. The results are broadly consistent, though there are small differences consistent with changes to the waveform model. Finally, in blue we show the recovery with \nsbh. The NSBH model has a slight preference for less equal mass ratio. Note that we have reweighted the samples by dividing by the prior on $\tilde{\Lambda}$, as done in Ref.~\cite{LIGOScientific:2018mvr}.}
    \label{fig:GW170817-PE}
\end{figure}

First, we obtain that the median-recovered matched-filter SNR for each
  waveform model is $32.7$. Since the \nsbh\ and \bns\ models recover the signal
  with a similar SNR, we do not find a clear preference either for a
  NSBH or BNS signal, when we only consider the GW data. Moreover, in
  Fig.~\ref{fig:GW170817-PE}, we show the recovery of the mass ratio
  and tidal deformability $\tilde{\Lambda}$ which is given by
  \begin{equation}
  \tilde{\Lambda} = \frac{16}{13}\frac{(m_1+12m_2)m_1^4\Lambda_1 + (m_2+12 m_1) m_2^4 \Lambda_2}{(m_1+m_2)^5}.
  \label{eq:LamTilde}
  \end{equation}
   In order to more easily
  compare with results in Ref.~\cite{LIGOScientific:2018mvr}, we show the mass
  ratio $q\equiv Q^{-1}=M_{\rm NS}/M_{\rm BH}$. There is a preference
  for unequal mass ratios in the \nsbh\  case, due to tidal disruption that 
  occurs for higher mass ratios. Since $\tilde{\Lambda}$ depends non-trivially on
  the mass ratio and individual tidal parameters, and since $\Lambda_1$ is fixed to zero in 
 the prior for \nsbh\ but not for the BNS models, the priors on $\tilde{\Lambda}$ for the BNS and NSBH
  models are quite different. In order to make a fair comparison between
  the posteriors on $\tilde{\Lambda}$, we divide each posterior by the prior on $\tilde{\Lambda}$,
   effectively obtaining a flat prior on $\tilde{\Lambda}$, as was done in Ref.~\cite{LIGOScientific:2018mvr}.
   We give the median and 90\%
  credible intervals for the masses, $\chi_{\rm eff}$, $\tilde{\Lambda}$, and matched filter network SNR, in Table~\ref{tab:GW170817-numbers}.

\begin{table*}
\begin{tabular}{c||c|c|c|c|c}
  \hline
  \hline
Approximant & $m_1/M_\odot$ & $m_2/M_\odot$ & $\chi_{\rm eff} / 10^{-3}$ & $\tilde{\Lambda}$ & Matched Filter SNR  \\
\hline
\bns\ (GWTC-1 version) & $1.47^{+0.15}_{-0.09} $ & $1.28^{+0.08}_{-0.11} $ &  $3.4^{+14}_{-8}  $ & $456^{+565}_{-307}$ & $32.7^{+0.1}_{-0.1}$ \\
\bns\ (current version) & $1.47^{+0.15}_{-0.09} $ & $1.28^{+0.08}_{-0.11} $ & $3.6^{+14}_{-8} $ & $345^{+545}_{-217}$  & $32.7^{+0.1}_{-0.1}$ \\
\nsbh & $1.50^{+0.16}_{-0.11}  $ & $1.26^{+0.10}_{-0.11} $ & $3.4^{+17}_{-9} $ &  $301^{+518}_{-189}$ & $32.7^{+0.1}_{-0.1}$\\
  \hline
\end{tabular}
\caption{Median value and 90\% credible region for parameters during a reanalysis of GW170817 with \nsbh\ and \bns. The 3 PE runs give consistent results without a strong preference for BNS or NSBH based on GW data alone. The NSBH recovery appears to have a slight preference for unequal mass ratios and a smaller effective tidal deformability $\tilde{\Lambda}$ compared to the BNS recoveries.}
\label{tab:GW170817-numbers}
\end{table*}

\section{Conclusion}

\label{sec:Conclusions}

In this work we have built an aligned-spin NSBH waveform model 
based on the EOB framework \cite{PhysRevD.59.084006,PhysRevD.62.064015,Damour:2000we}, the NRTidal approach \cite{Dietrich:2017aum,Dietrich:2019kaq} and  NR simulations \cite{PhysRevD.88.064017,Foucart:2018lhe,Foucart:2014nda,Kyutoku:2010zd,Kyutoku:2011vz}: \nsbh. In building the model, we have used final mass and spin fits from Ref.~\cite{Zappa:2019ntl},  the $\Lambda_{\rm NS}-C_{\rm NS}$ relations of Ref.~\cite{Yagi:2016bkt}, and fits for the disk mass from Ref.~\cite{Foucart:2012nc}. This model
incorporates a suitable tapering of the frequency-domain waveform's amplitude in regions where tidal disruption
occurs building on the amplitude model of Ref.~\cite{Pannarale:2015jka}, as evinced from physical considerations of tidal effects 
as the NS plunges into the BH, and from NR simulations of those sources. 
Tidal corrections to the frequency-domain waveform's phase have been computed using the
NRTidal framework.
We have shown that \nsbh\ gives good agreement with NR simulations by comparing the waveforms in the
 frequency domain (Fig.~\ref{fig:SXS-FD}) and time domain (Fig.~\ref{fig:SXS-TD}), as well as by computing
  the unfaithfulnesses shown in Tables~\ref{tab:SXS_comparisons} and ~\ref{tab:SXS_hybrid_comparisons}. 
  In Fig.~\ref{fig:SXS-TD-2}, we compare the model with two new, highly accurate, simulations from the SXS Collaboration  
  of disruptive NSBHs with highly spinning BHs, which we used for validation. We find very good agreement across a large 
  number of cycles. We also performed the same comparisons with the recently published model PhenomNSBH, and find similar levels of agreement with NR.
 In Figs.~\ref{fig:Injection-1} and~\ref{fig:Injection-2}, we have demonstrated that the model can be used to
infer properties of NSBH systems using software injections, and that at large enough SNR assuming the wrong source class can lead to biased astrophysical inferences. Finally, we have reanalyzed GW170817 with the
hypothesis that it is a NSBH  instead of a  BNS. In Table~\ref{tab:GW170817-numbers}, we see the results are broadly consistent, although there seems to be a slight preference for smaller tidal deformability and unequal masses when recovering with \nsbh. 

In the future, we plan to extend and improve the \nsbh\ waveform model in various ways. 
A relatively simple, but important extension is to incorporate information from 
modes beyond the quadrupolar one using SEOBNRv4HM~\cite{Cotesta:2018fcv} as a baseline.
This  is particularly relevant, since the NS can be tidally disrupted also in cases  
in which the mass ratio is larger than one and the BH spin is large. Another crucial 
improvement is to extend the model to precessing NSBH binaries, since some astrophysical 
scenarios predict that the BH spin may be misaligned with respect to the orbital angular momentum. 
As more high quality NR simulations of NSBHs become available, it will also be possible
to develop a more accurate model for the transition from disruptive to non-disruptive mergers.
It will also be interesting to study the effect of using different tidal models in order to 
quantify uncertainty in the tidal part of the waveform.
Finally, as we have mentioned, there is currently no model that smoothly covers the 
full range of source classes: BBH, NSBH, and BNS. Building a model which can capture all 
of relevant physics is an important future goal.

\section*{Acknowledgments}

We would like to thank Koutarou Kyutoku and Masaru Shibata for
providing us with the numerical-relativity waveforms from the \verb+SACRA+ code. We would like to
thank Frank Ohme, Jonathan Thompson, Edward Fauchon-Jones, and
Shrobana Ghosh for reviewing the LAL implementation of the 
\nsbh\ waveform model. We are grateful to Katerina Chatziioannou for comments on the manuscript, and Luca Prudenzi for useful discussions. 

TD acknowledges support by the European
Union’s Horizon 2020 research and innovation program under grant
agreement No. 749145, BNS mergers.
TH acknowledges support from NWO Projectruimte grant GW-EM NS and the DeltaITP, and thanks the YKIS 2019.
F.F. gratefully acknowledges support from the NSF through grant PHY-1806278. M.D gratefully acknowledges support from the NSF through grant PHY-1806207.  H.P. gratefully acknowledges support from the NSERC Canada. L.K. acknowledges support from NSF grant PHY-1606654 and PHY-1912081. M.S. acknowledge support from NSF Grants PHY170212 and PHY-1708213. L.K. and M.S. also thank the Sherman Fairchild Foundation for their support.

 Computations for the review were
done on the \texttt{Hawk} high-performance compute (HPC) cluster at 
Cardiff University, which is funded by STFC grant ST/I006285/1. Other computations for this work
were done on the HPC clusters \texttt{Hypatia} at the Max Planck Institute 
for Gravitational Physics in Potsdam, and at \texttt{CIT} at Caltech, 
funded by National Science Foundation Grants PHY-0757058 and PHY-0823459. We extensively used the \texttt{numpy} \cite{numpy}, \texttt{scipy}
\cite{2020SciPy-NMeth}, and \texttt{matplotlib} \cite{Hunter:2007}
libraries. 

This research has made use of data, software and/or web
tools obtained from the Gravitational Wave Open Science Center
(https://www.gw-openscience.org), a service of LIGO Laboratory, the
LIGO Scientific Collaboration and the Virgo Collaboration. LIGO is
funded by the U.S. National Science Foundation. Virgo is funded by the
French Centre National de Recherche Scientifique (CNRS), the Italian
Istituto Nazionale della Fisica Nucleare (INFN) and the Dutch Nikhef,
with contributions by Polish and Hungarian institutes.

\appendix

\section{Explicit form of amplitude correction}
\label{app:w_definition}

The amplitude corrections are parameterized based on the model presented in Ref.~\cite{Pannarale:2015jka}. We use the same parametric form for each component of the amplitude correction, and refit the coefficients. We have streamlined the notation.

\subsection{Non-disruptive}
The non-disruptive window function given in Eq.~\ref{eq:wND} contains the parameters $f_{\rm ND}$, $\sigma_{\rm ND}$, and $\epsilon_{\rm ND}$, which we compute as
\begin{subequations}
\begin{eqnarray}
   f_{\rm ND} &=& f_{\rm RD}, \\
   \sigma_{\rm ND} &=& \bar{\sigma}_{\rm ND}+2 w^-(x;x_0,\sigma_x),    \\
   \epsilon_{\rm ND} &=& w^+(y;y_0,\sigma_y), 
 \end{eqnarray}
 \end{subequations}
 where $f_{\rm RD}$ is the ringdown frequency, which we estimate in terms of the final mass and spin using the fits from Ref.~\cite{Berti:2005ys}.
 Following Refs.~\cite{Pannarale:2013uoa,Pannarale:2015jka}, we have introduced $x$ and $y$, which are a measure of how close the merger is to becoming disruptive. These quantities appear inside of window functions in order to ensure that the corrections to the ringdown are smoothly turned off ($\epsilon_{\rm ND}\rightarrow 0, \sigma_{\rm ND}\rightarrow \infty$) as the merger becomes less disruptive and therefore more like a BBH. For large $\sigma_{\rm ND}$ and $\epsilon_{\rm ND}$ on the intrinsic parameters of the binary. are determined by the tidal frequency, ringdown frequency, NS compactness, and BH spin via
\begin{subequations}
 \begin{eqnarray}
   x &=& \left(\frac{f_{\rm RD}-f_{\rm tide}}{f_{RD}}\right)^2 + x_C C_{\rm NS} + x_\chi \chi_{\rm BH},  \\
   y&=& \left(\frac{f_{\rm RD}-f_{\rm tide}}{f_{RD}}\right)^2 + y_C C_{\rm NS} + y_\chi \chi_{\rm BH}.
\end{eqnarray}
\end{subequations}
The nine coefficients $\{\bar{\sigma}_{\rm ND},x_0,\sigma_x,x_C,x_\chi,y_0,\sigma_y,y_C,y_\chi\}$ were determined by a fitting procedure as described in Sec.~\ref{sec:fits}. Their values are given in Table~\ref{tab:w_parameters}.

\subsection{Disruptive}

The disruptive window correction in Eq.~\ref{eq:wD} is defined in terms of $f_{\rm D}$ and $\sigma_{\rm D}$, which we parameterize as
\begin{subequations}
\begin{eqnarray}
   f_{\rm D} &=& \Big( a_0 + a_M \frac{M_{\rm b,torus}}{M_{\rm b,NS}} +   a_C C_{\rm NS} + a_\nu \sqrt{\nu} + a_\chi \chi \Big) f_{\rm tide}  \nonumber \\
   \\
      \sigma_{\rm D} &=& b_0+ b_M \frac{M_{\rm b,torus}}{M_{\rm b,NS}} + b_C C_{\rm NS} + b_\nu \sqrt{\nu} + \sum_{k=1}^3 b^{(k)}_\chi \chi^k. \nonumber  \\
   \end{eqnarray}
   \end{subequations}
In this expression, there are twelve coefficients $\{a_0,a_M,a_C,a_\nu,a_\chi,b_0,b_M,b_C,b_\nu,b_\chi^{(1)},b_{\chi}^{(2)},b_{\chi}^{(3)}\}$ which were determined by a fitting procedure as described in Sec.~\ref{sec:fits}. Their values are given in Table \ref{tab:w_parameters}. As in the case of Ref.~\cite{Pannarale:2015jka}, we find that the parameters $b_\chi^{(1)},b_{\chi}^{(2)},b_{\chi}^{(3)}$ do not decrease monotonically.

\begin{table}[ht]
\centering
\begin{tabular}{c|c}
  \hline
  \hline 
Parameter  &  Value \\ 
 \hline
   $\bar{\sigma}_{\rm ND}$ & \sigmaNDone \\
   $x_0$    & \xbar \\
  $\sigma_x$ & \sigmax \\   
  $x_{C}$    & \xC \\
  $x_{\chi}$    & \xchi \\
  $y_0$    & \ybar \\
  $\sigma_y$ & \sigmay \\  
  $y_{C}$ & \yC \\
  $y_{\chi}$ & \ychi \\
  \hline
  $a_0$    & \azero \\
  $a_M$    & \aM \\
  $a_C$ & \aC \\
  $a_\nu$ & \anu \\
  $a_\chi$ & \achi \\
  $b_0$    & \bzero \\
  $b_M$    & \bM \\
  $b_C$ & \bC \\
  $b_\nu$ & \bnu \\
  $b_{\chi}^{(1)}$ & \bchione \\
  $b_{\chi}^{(2)}$ & \bchitwo \\
  $b_{\chi}^{(3)}$ & \bchithree \\
  \hline
\end{tabular}
    \caption{Parameters for the amplitude correction $w_{\rm corr}(f)$. The parameters above the line appear in the correction for tidally disruptive mergers, the parameters below the line appear in the correction for disruptive mergers.}
    \label{tab:w_parameters}
\end{table}

\bibliography{refs}

\end{document}

%% file: macros.tex
\def\azero{1.27280}
\def\aM{-1.68735}
\def\aC{-1.43369}
\def\anu{-0.510033}
\def\achi{0.280002}
\def\bzero{0.185326}
\def\bM{-0.253476}
\def\bC{0.251061}
\def\bnu{-0.284595}
\def\bchione{-0.000757100}
\def\bchitwo{0.018089075}
\def\bchithree{0.028545184}

\def\sigmaNDone{0.0225006}
\def\xbar{-0.0923660}
\def\sigmax{0.0187155}
\def\xC{-0.486533}
\def\xchi{-0.0314394}
\def\ybar{-0.177393}
\def\yC{0.493376}
\def\ychi{0.0569155}
\def\sigmay{0.771910}

\def\bbh{SEOBNR\_BBH}
\def\bns{SEOBNR\_BNS}
\def\nsbh{SEOBNR\_NSBH}